\documentclass[sigconf]{acmart}

\renewcommand\footnotetextcopyrightpermission[1]{} 
\makeatletter
\renewcommand\@formatdoi[1]{\ignorespaces}
\makeatother
\acmConference[]{}{}{}

\AtBeginDocument{%
  \providecommand\BibTeX{{%
    \normalfont B\kern-0.5em{\scshape i\kern-0.25em b}\kern-0.8em\TeX}}}

\begin{document}


\title{Becoming the Super Turker:\\Increasing Wages via a Strategy from High Earning Workers}
\author{Saiph Savage{$^1,^2$}}
\affiliation{%
  \institution{Universidad Nacional Autonoma de Mexico (UNAM){$^1$}}}
\email{saiph.savage@mail.wvu.edu}

\author{Chun Chiang{$^2$}, Susumu Saito{$^3$}}
\affiliation{%
  \institution{West Virginia University (WVU){$^2$}}
  \institution{Waseda University{$^3$}}
}

\author{Carlos Toxtli{$^2$}, Jeffrey Bigham{$^{4}$}}
\affiliation{%
 \institution{Carnegie Mellon University (CMU){$^4$}}
 }

\renewcommand{\shortauthors}{Saiph Savage, et al.}

\begin{abstract}
Crowd markets have traditionally limited workers by not providing transparency information concerning which tasks pay fairly or which requesters are unreliable. Researchers believe that a key reason why crowd workers earn low wages is due to this lack of transparency. As a result, tools have been developed to provide more transparency within crowd markets to help workers. However, while most workers use these tools, they still earn less than minimum wage. We argue that the missing element is guidance on how to use transparency information. In this paper, we explore how novice workers can improve their earnings by following the transparency criteria of Super Turkers, i.e., crowd workers who earn higher salaries on Amazon Mechanical Turk (MTurk). We believe that Super Turkers have developed effective processes for using transparency information. Therefore, by having novices follow a Super Turker criteria (one that is simple and popular among Super Turkers), we can help novices increase their wages. For this purpose, we: {\em (i)} conducted a survey and data analysis to computationally identify a simple yet common criteria that Super Turkers use for handling transparency tools; {\em (ii)} deployed a two-week field experiment with novices who followed this Super Turker criteria to find better work on MTurk. Novices in our study viewed over 25,000 tasks by 1,394 requesters. We found that novices who utilized this Super Turkers' criteria earned better wages than other novices. Our results highlight that tool development to support crowd workers should be paired with educational opportunities that teach workers how to effectively use the tools and their related metrics (e.g., transparency values). We finish with design recommendations for empowering crowd workers to earn higher salaries.
\end{abstract}
\maketitle

\section{Introduction}
Amazon Mechanical Turk (MTurk) is the most popular crowd market \cite{kuek2015global}. 
It allows crowd workers (Turkers) to earn money from micro jobs involving human-intelligence-tasks (HITs). Although MTurk brings new jobs to the economy, most Turkers still struggle to earn the U.S. minimum wage (\$7.25) \cite{hara2018data,berg2018digital}. This is problematic considering that ``earning good wages'' is the primary motivator of crowd workers  \cite{berg2015income,kaufmann2011more,berg2018digital}. Many believe that the lack of transparency on MTurk is the root cause why Turkers are not being fairly compensated \cite{hara2018data}. Economists consider that a market is transparent when all actors can access vast information about the market such as the products, services, or capital assets \cite{strathern2000tyranny}. Similarly, Silberman et al. discusses how the lack of transparency on gig markets affects workers earnings: ``A wide range of processes that shape platform-based workers' ability to find work and receive payment for work completed are, on many platforms, opaque \cite{metall2016frankfurt}.'' MTurk has primarily focused on providing transparency information solely to requesters by allowing them to access in-depth knowledge about Turkers. However, MTurk has traditionally provided more limited information to workers (e.g., Turkers previously could not profit from knowledge about requesters' previous hiring record or the estimated hourly wage of the tasks on the market, although as of July 2019, this has started to change\footnote{\url{https://blog.mturk.com/new-feature-for-the-mturk-marketplace-aaa0bd520e5b}}). This lack of transparency for workers can lead them to invest significant time in a task but receive anywhere from inadequate to no compensation.

To begin addressing the issue of transparency, scholars and practitioners have developed web browser extensions \cite{irani2013turkopticon,turkerview} or created online forums\footnote{\url{www.turkernation.com/}} to bring greater transparency to Turkers. These tools and forums provide Turkers with otherwise unavailable information about requesters, tasks, and expected payment. For instance, TurkerView allows workers to obtain an overview of the expected hourly wage they would receive if they worked for a particular requester. However, while an ever increasing number of workers are using these tools for transparency  \cite{kaplan2018striving}, only a fraction of Turkers' earnings are well above the minimum wage \cite{hara2018data}. The problem is that utilizing transparency tools to earn higher wages is not straightforward. 
Each transparency tool displays several different metrics. This poses the question, which metric should a Turker use to ensure better wages? This complexity has likely led most Turkers to employ transparency tools ineffectively \cite{kaplan2018striving,saito2019turkscanner}.

Despite the challenges associated with using transparency tools, top earning crowd workers on MTurk, those that make above the minimum wage, have emerged. These rare workers are commonly referred to as ``Super Turkers'' because they earn ``superior'' wages. They have earned this name despite market conditions such as limited availability of tasks and a greater percentage of low-paying tasks\cite{berg2018digital,hara2018data,bohannon2011social}. We argue that it is Super Turkers' ability to use transparency that brings them a unique advantage\cite{bohanec2009decision}. Our goal was to uncover one of the ways in which Super Turkers used transparency, and then guide novices to follow that same criteria. 

For this purpose, we first questioned Super Turkers on how they used specific transparency tools to decide which tasks to perform. We then conducted a data analysis of their responses to computationally identify a Super Turker criteria that denoted one of the ways in which Super Turkers decided to use transparency. We rooted our data analysis on the ``Analytic Hierarchy Process'', a well established multi-criteria decision-making approach \cite{vaidya2006analytic,forman1990multi,kasperczyk1996analytic}. By using the Analytic Hierarchy Process we identify in a hierarchical form the level of importance that Super Turkers give to different transparency metrics when deciding what tasks to perform. We utilized a hierarchical process given that transparency information is not always available and workers have limited time to decide what tasks to perform (especially as time spent finding labor is time where workers are not paid \cite{gray2019ghost}). We argue that the hierarchy helps workers to more rapidly identify what transparency metrics they should analyze; and if any of those metrics are unavailable, they have a plan of what else to rapidly inspect \cite{casler2013separate}. We also aimed for this criteria to be simple and popular among Super Turkers. Simplicity was important to make it easier for novices to follow \cite{payne1996time}. Popularity was important to have a decision criteria that was representative of Super Turkers (although it is possible and likely that Super Turkers also have other more complex criteria for deciding how to use transparency.)

Once we had an identified transparency criteria that Super Turkers utilized to select work, we conducted a field experiment to investigate how the hourly wage of novices changed when following such criteria. Notice that it is not simple to run a field experiment that can track how workers actually increment their hourly wages. MTurk does not provide any information about the hourly wage of a particular task nor how much time it would take workers to complete a given HIT. It is, therefore, not straightforward 
how one might calculate the change in workers' wages over time \cite{hara2018data}. To overcome these challenges, we developed a plugin that logs workers' behavior, calculates how much time workers spend on each task, and estimates each worker's hourly wage per HIT\footnote{Since calculating hourly wage of Turkers has proven a difficult task for researchers in the past, we have released the plugin to help other researchers:\\
\url{https://research.hcilab.ml/superturker}}. The plugin is inspired on prior research and related tools\cite{callison2014crowd}. 

Equipped with our plugin and the Super Turker criteria, we ran a two-week field experiment. We had  real-world novices perform over 25,000 tasks on MTurk by 1,394 requesters, with the experimental group of workers following the Super Turker criteria, and the control group not receiving any additional guidance. Our study uncovered that having novices follow the Super Turker criteria did empower them to increase their income. We finish with design recommendations for tools and platforms to increase crowd workers' wages. We advocate for tools that bring transparency, and also teach workers how to best make use of that transparency.

\section{Related Work}

\subsection{Work Environment on Crowd Markets}

Crowdsourcing not only facilitates the generation of ground truth for machine learning \cite{deng2009imagenet}, but  also enables novel crowd-powered technology \cite{bigham2010vizwiz,huang2018evorus}. Technology companies use crowdsourcing as ``ghost work'', which is unperceived by end users \cite{gray2019ghost,martin2014being}.
However, criticism surrounding crowd markets compares them to  sweatshops or ''markets for lemons''\cite{vakharia2015beyond,ipeirotis2010mechanical,cushing2012dawn,silberman2009fifteen}. 
Receiving wages that are less than the U.S. hourly minimum wage, which is \$7.25 USD, is one of the most significant disadvantages for workers in crowd markets \cite{irani2013turkopticon,irani2016stories,KatzAmazon,kasunic2019crowd,salehi2015we,bergvall2014mazon,international2016non,horton2011condition,horton2010labor,hitlin2016research,hara2018data}.
Besides the negative factor of low wages, requesters create tasks for workers to perform, but are able to arbitrarily reject the submitted work once the labor has been accomplished \cite{gray2019ghost}.

Additionally, crowd workers spend a significant amount of time performing invisible and unpaid labor, e.g., acquiring tasks, learning how to perform assigned tasks, and resolving conflicts with the platform or requesters when discrepancies concerning payment occur \cite{hara2018data,sannon2019privacy,han2019all,gadiraju2017modus}. One of the main reasons for this is that crowd markets have imposed transaction costs, which were traditionally assumed by companies, onto workers \cite{de2015rise,gray2019ghost}. Transaction costs are the expenses associated with managing the exchange of goods or services. Researchers have coined this situation  ``algorithmic cruelty'' as the algorithms behind the crowd market are generating critical pain points for workers, such as having no recourse if their account becomes unfairly blocked, if their completed work is arbitrarily rejected, or if they are not fairly compensated. 

To further complicate the situation, crowd markets do not provide the same information to workers than to requesters. Usually, requesters are granted access to a large amount of information concerning the events in the marketplace; while workers have a much more limited perspective  \cite{irani2013turkopticon,irani2016stories}. For example, MTurk allows requesters to view the previous performances and interactions that workers have had on the platform \cite{hara2018data}; while workers can only discern very little about what requesters have done previously (e.g., amount of rejected work, amount of unfairly paid tasks, or whether they are frauds\cite{irani2013turkopticon,gadiraju2019understanding}). 

A consequence of this limited information (coined a ``lack of transparency'' by researchers) is that crowd workers struggle to find fairly paid tasks or even be paid. Additionally, crowd workers lack basic benefits, e.g., paid sick leave, time off, and health insurance \cite{harmon2018rating,gray2019ghost}.
Also, the work on crowdsourcing platforms typically does not help workers to advance their career
\cite{kasunic2019crowd}.
As crowd markets continue to grow, they threaten the hard earned workers' rights attained through the labor movements \cite{harmon2018rating}. Additionally, labor market oversight is more difficult in a crowd market economy\cite{kassi2018online}.

\subsection{Tools for Crowd Market Transparency}

\begin{figure}
\centering
  \includegraphics[width=.95\columnwidth]{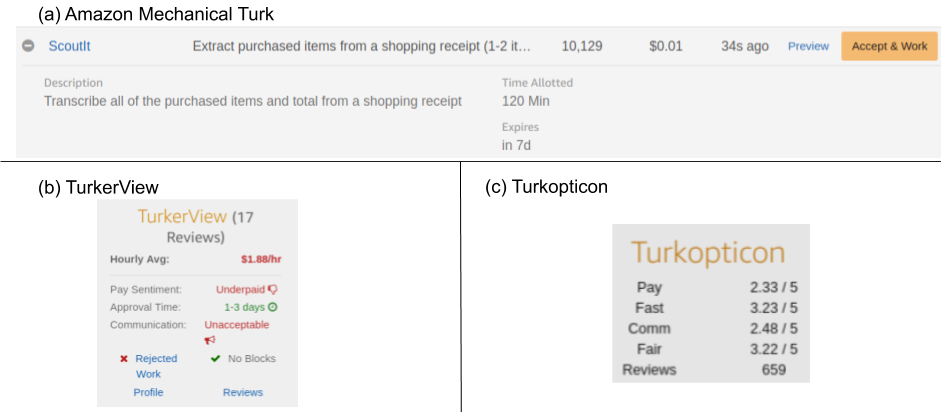}
  \vspace{-2mm}
  \caption{What a turker sees: (a) the HITs, (b) requester's reputation metrics on TurkerView, and (c) requester's reputation metric on Turkopticon.}
  \label{fig:interface}
\end{figure}

To begin addressing the unfairness that crowd workers experience, researchers have created tools that bring more transparency to the crowd market (e.g., tools that help workers better comprehend information about requesters and the market in general.) In this paper we refer to these tools as ``transparency tools.'' These approaches believe that through transparency workers can learn how to avoid unreliable requesters and earn better wages \cite{mcinnis2016taking}. Researchers and practitioners have developed different forums and browser extensions to help workers measure the reputation of requesters (e.g., how they previously interacted with workers) \cite{mcinnis2016taking}. Crowd workers use Turkopticon \cite{irani2013turkopticon} and TurkerView \cite{turkerview} to evaluate requesters \cite{sannon2019privacy,kaplan2018striving}. Fig. \ref{fig:interface} displays the interface used on the Mturk, Turkopticon, and TurkerView.
Turkopticon is an opensource tool that allows crowd workers to rate the requesters with 4 ``attributes'' in a 5 point Likert-scale: generosity (``pay''), promptness (``fast''), communicativity (``comm'') and fairness (``fair'') \cite{irani2013turkopticon}. Crowd workers can also leave text descriptions to illustrate each requester and how they interact with workers, as well as the type of tasks they post on the platform. TurkerView is another transparency tool that allows workers to visualize requesters' reputations and offers metrics that are similar to Turkopticon.  TurkerView does have some differences. First, it is not opensource. The tool has focused on commercializing its intelligent algorithms that predict how a requester will behave based on her interactions with workers who are using Turkerview\cite{turkerview}. Turkerview also offers a metric that predicts the hourly wage that a given requester is likely to pay (which is not exact per task but provides an overall picture of how that requester operates). Browser extension tools and forums have increased transparency on MTurk. However, despite the availability of such tools, novices still fail to recognize which requester will pay fairly \cite{kaplan2018striving,hara2018data}. (These labor conditions might change in the future with the introduction of ``One Line Of Code'' to automatically ensure fair wages\cite{whiting2019fair}. But these approaches depend on the requester and the platforms wanting to be ``fair'', which is not always the case\cite{gray2019ghost}). 

Together, this suggests that unguided transparency is not sufficient to ensure higher wages. We argue that adopting a Super Turker criteria in a Analytical Hierarchy Process approach offers the necessary guidance to novices to utilize transparency tools to earn higher wages.
Our study aims to: 
\begin{itemize}
\item identify a common and simple criteria that Super Turkers employ for using transparency tools to find work
\item guide novices to follow the identified criteria
\item increase novices' earning potential
\end{itemize}

\section{Uncovering Super Turker Practices}
Our goal was to identify a common set of criteria that Super Turkers implemented when using transparency tools. Each Super Turker might value many different criteria. However, understanding that novices progress to experts through repetition\cite{gadiraju2015training,persky2017moving}, we were interested in identifying criteria that were simple and widely accepted, so that novices could easily apply the criteria to earn higher wages. For this purpose, we created a survey that questioned Super Turkers on their criteria for using transparency tools, and then conducted a data analysis over the responses to identify a simple yet popular criteria that novices could easily follow.
\subsection{Survey: Method} 
The survey contained 18 required questions and took 3 to 5 minutes to complete. Our survey was rooted in prior research \cite{ross2010crowdworkers} and based on our research questions. Participants were paid \$0.60 USD according to a legal hourly wage of \$7.25 USD. The survey was only available to workers, deemed Super Turkers, i.e., workers who had done over 10,000 HITs and who earned more than the U.S minimum wage.
\begin{figure}
  \begin{center}
    \includegraphics[width=.95\columnwidth]{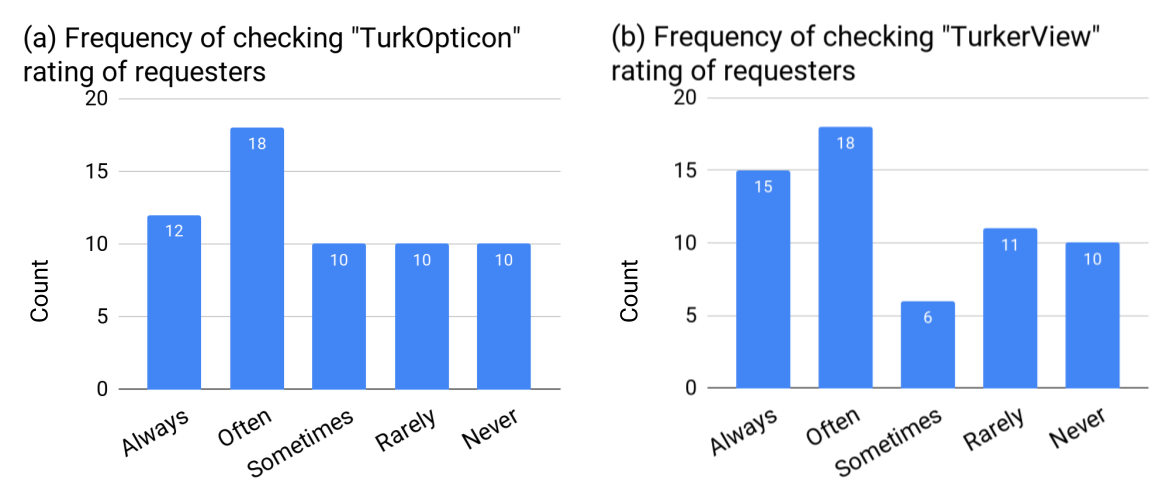}
  \end{center}
  \vspace{-2mm}
  \caption{ Frequency of how often Super Turkers checked requesters' ratings on Turkopticon and TurkerView. }
\label{fig:reputation}
\end{figure}
Similar to \cite{ross2010crowdworkers,kaplan2018striving}, our survey began by asking Super Turkers demographic information and questions about their MTurk work experiences. These questions asked crowd workers about their weekly working hours and how long they had been on MTurk. Next, we asked workers to create flowcharts by selecting and sorting through a list of steps. They denoted the order in which they used different transparency tools and the metrics they analyzed from each tool.
For validity purposes, we based the list on prior work that has studied workers' actions around HITs \cite{yuen2012task,saito2019turkscanner} and the metrics that transparency tools, e.g., Turkopticon or TurkerView, share with workers \cite{turkerview,irani2013turkopticon}. Participants could either select and sort all steps on the list or select and sort the specific few steps that were key to them. After this question, each Super Turker had a flowchart denoting her process for using transparency tools to select HITs. Next, we studied the importance that Super Turkers gave to different transparency tools and their associated metrics. For each type of transparency tool that Super Turkers stated that they used in their flowchart, the survey asked them how they handled such information. The questions included how often they checked the information and how important the particular metrics associated with the information were. We questioned Super Turkers about the minimum acceptable scores they had for each metric. We used 5-point Likert scale questions to ask Super Turkers about how often they checked the metric (frequency) and the importance they gave to the metric when making decisions. The options in our Likert scales questions were based on the anchors created by Vagias \cite{vagias2006likert}. The minimum acceptable score questions were slider questions that ranged between scores 1 and 5 or ``not applicable.'' 

 
\subsection{Survey: Findings}
100 Super Turkers completed the survey. To avoid malicious or distracted workers, we added two attention check questions into the survey. This resulted in us keeping 68 responses for the analysis, the other responses were excluded because of failed attention checks.

\subsubsection{Understanding Super Turkers' Key Transparency Metrics}


88\% of the Super Turkers that participated in our survey stated that they used transparency tools. 
Fig. \ref{fig:reputation} presents an overview of how frequently Super Turkers used each of these tools to access transparency information about requesters. Super Turkers primarily used both TurkerView and Turkopticon plugins to evaluate requesters, although TurkerView was used slightly more frequently than Turkopticon. Next, we analyzed the type of metrics that Super Turkers took into account when using these tools.

\begin{figure}
\centering
  \includegraphics[width=.99\columnwidth]{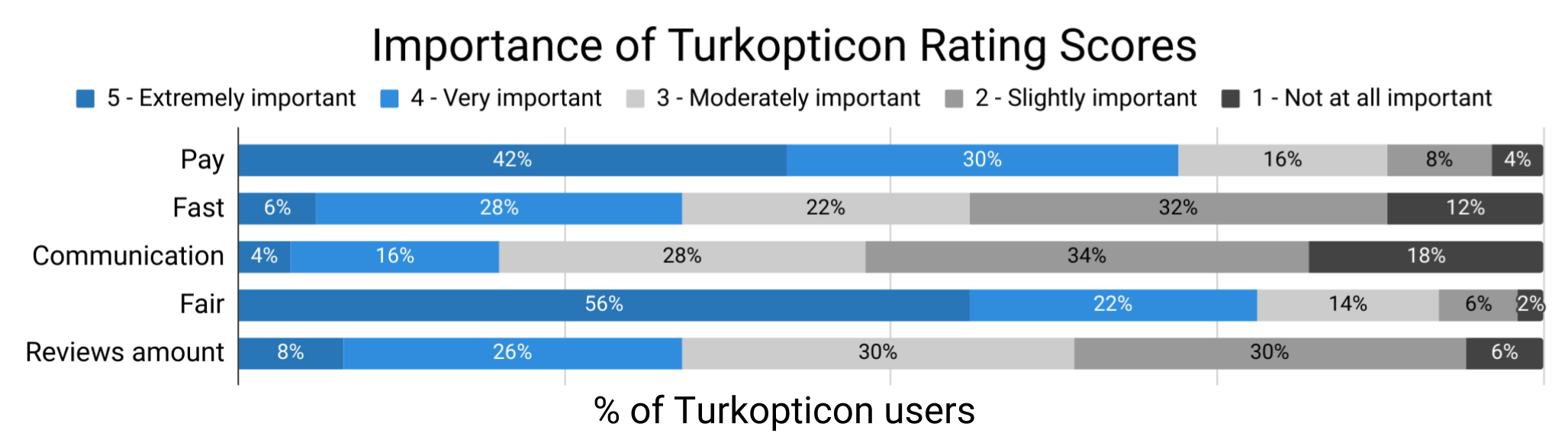}
  \vspace{-2mm}
  \caption{Overview of the importance that Super Turkers gave to different  ``Turkopticon'' metrics about requesters: \textit{fair} was the most important, followed by \textit{pay}.
  }
  \label{fig:Importance(Turkopticon)}
\end{figure}

Fig. \ref{fig:Importance(Turkopticon)} presents an overview of how important each Turkopticon metric was for Super Turkers. The most important metrics were \textit{TO\_fair} and \textit{TO\_pay} (Notice that we use the term ``\textit{TO}'' to distinguish Turkopticon metrics from TurkerView ``\textit{TV}'' metrics). More than half of the respondents (78\%) considered that the \textit{TO\_fair} metric was either ``5 - extremely important'' or ``4 - very important'' when selecting HITs. Meanwhile, 72\% of the respondents deemed the \textit{TO\_pay} metric as  ``5- extremely important'' and ``4 - very important.'' \textit{TO\_fair} describes whether the requester rejects or approves the work in a fair way, and \textit{TO\_pay} metric represents how well the requester pays. On Mturk, requesters can gratuitously reject workers and can indiscriminately refuse payment upon completion of the tasks. Thus, this metric measures how likely is an individual requester to reject or accept work in reasonable (fair) manner. 

After understanding how important these different transparency metrics were to Super Turkers, we sought to understand the values that Super Turkers expected or looked for in each metric. Recognizing these thresholds can help novice workers to better utilize the metrics and be more effective at finding fair HITs. Our results indicated that Super Turkers had different expectations for each metric. 92\% of our Super Turkers who used Turkopticon expressed that they had a basic requirement score for \textit{TO\_fair} (e.g., they only considered HITs who had a \textit{TO\_fair} score above a certain threshold), and 90\% of the Super Turkers had a basic requirement score for \textit{TO\_pay}. The average requirement score of \textit{TO\_fair} was 3.69 (SD=0.78) and of \textit{TO\_pay} was 3.2 (SD=0.9).  


\begin{figure}
\centering
  \includegraphics[width=.99\columnwidth]{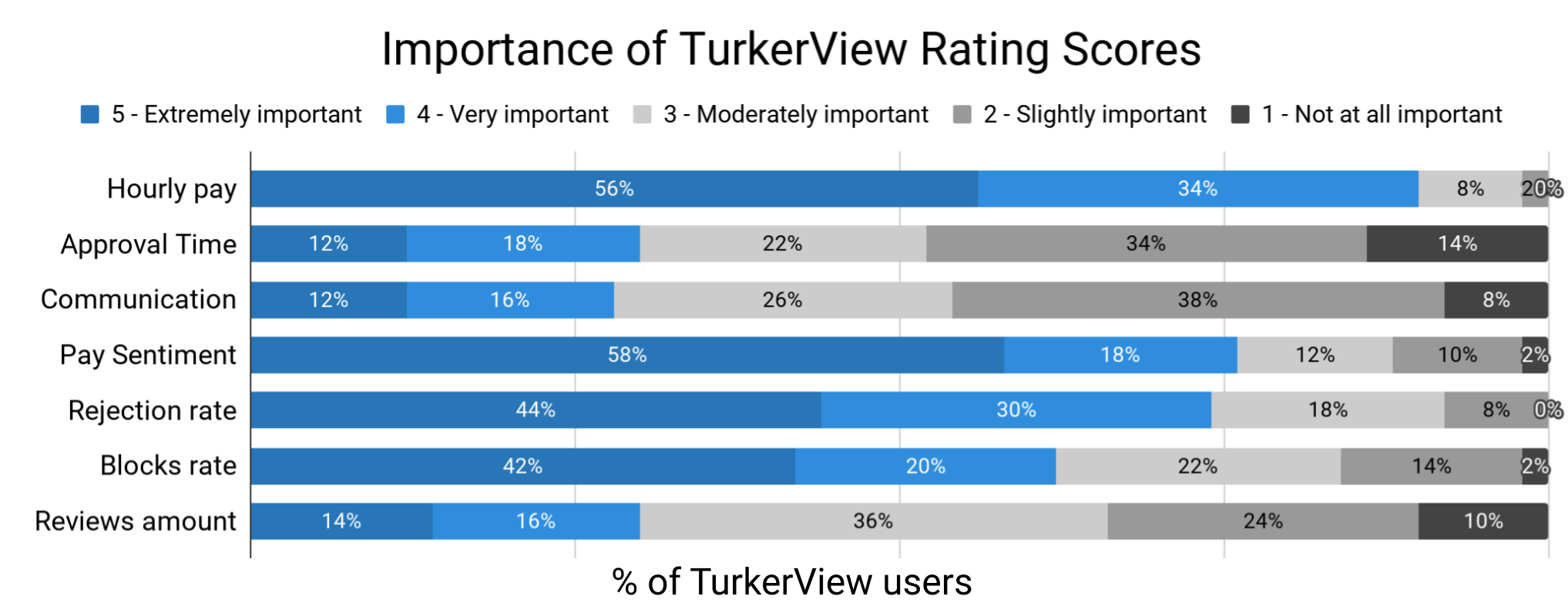}
  \vspace{-2mm}
  \caption{The importance of each ``TurkerView'' reputation attributes. \textit{Hourly pay} and \textit{Fair} are crucial when Super Turkers select a HIT to work.}
  \label{fig:Importance(TurkerView)}
\end{figure}



We conducted a similar analysis for TurkerView. Fig. \ref{fig:Importance(TurkerView)} shows that for Super Turkers, the most valuable transparency metric on TurkerView was: \textit{TV\_hourly\_pay}, followed closely by \textit{TV\_fair}, \textit{TV\_rejection\_rate} and \textit{TV\_block\_rate}. (Note the use of ``\textit{TV}'' in this case.) The \textit{TV\_hourly\_pay} metric estimates the hourly wage in USD that a requester will pay Turkers. This is based on previous workers' experiences performing tasks. 
To calculate this metric, TurkerView averages the hourly wage of each HIT posted by the requester and the amount of time it took workers to finish the task. \textit{TV\_fair} is a metric denoting whether workers believe that a requester pays fairly for the assigned task. \textit{TV\_rejection\_rate} and \textit{TV\_block\_rate} display the frequency that a particular requester rejects or blocks workers. When rejected or blocked by a requester, workers not only lose the remuneration for the completed work, but they also receive a bad record on MTurk. This hurts workers future employment opportunities. When analyzing the details of the metrics, we find that Super Turkers, on average, only accept the tasks of requesters whose \textit{TV\_hourly\_pay} was over \$8.29 USD/h (SD = 3.03).

However, we must also contend with the situation that TurkerView and Turkopticon do not hold information about all requesters on the platform. When this is the case, the primary metric to measure whether a HIT is worth doing is the reward it offers and the description of the task. From our survey, 82\% of Super Turkers had fundamental requirements on the metrics of the HIT rewards (i.e., for them to do a HIT, the reward needed to be above a certain threshold.) The minimum acceptable reward for a task averaged \$0.23 (SD = 0.23, median = 0.2). Note that Super Turkers took into account the overall reward rather than the hourly wage, because MTurk only provides information on rewards (how much the requester will pay in total if the worker completes the HIT). 

\subsubsection{Computing Super Turker Transparency Criteria.}\label{strategy}

Once we had an overview of the type of transparency metrics that Super Turkers considered, we wanted to identify the sequential order in which they evaluated these metrics. It is possible that Super Turkers have various sequences for how they use the available transparency metrics. Our goal was to identify criteria that was common (i.e.,used by many Super Turkers) and concise. Most important was a concise set of criteria for novices to efficiently and effectively implement. For this purpose, we took all the flowcharts that Super Turkers had generated, converted the steps into a text sequence, and used the longest common subsequence algorithm \cite{jiang1995approximation} to identify the criteria that was common and shortest among the Super Turkers in our study. The algorithm computed the following criteria:  

\begin{itemize}
    \item Work only with requesters whose:
    \begin{itemize}
    \item ``hourly pay'' on TurkerView is over \$8.29 USD/h (averaged from values that Super Turkers provided for this metric).
    \item If such transparency data is unavailable:
    \begin{itemize}
    \item work only with requesters whose ``fair''score on Turkopticon is over 3.69 (averaged).
     \end{itemize}
    \item If such transparency data is unavailable:
    \begin{itemize}
    \item perform tasks with reward > \$0.23 USD (averaged.)
    \end{itemize}
   \end{itemize}
\end{itemize}

Notice that the computed criteria defines a hierarchy of transparency metrics. The hierarchy helps in this case because not all transparency metrics are always available to workers, i.e., some can be missing. The criteria offers a way to potentially find ``good work'' 
even under labor market conditions of limited information.
\subsubsection{Super Turkers' Impressions of the Criteria.}
\begin{figure}
\centering
  \includegraphics[width=.50\columnwidth]{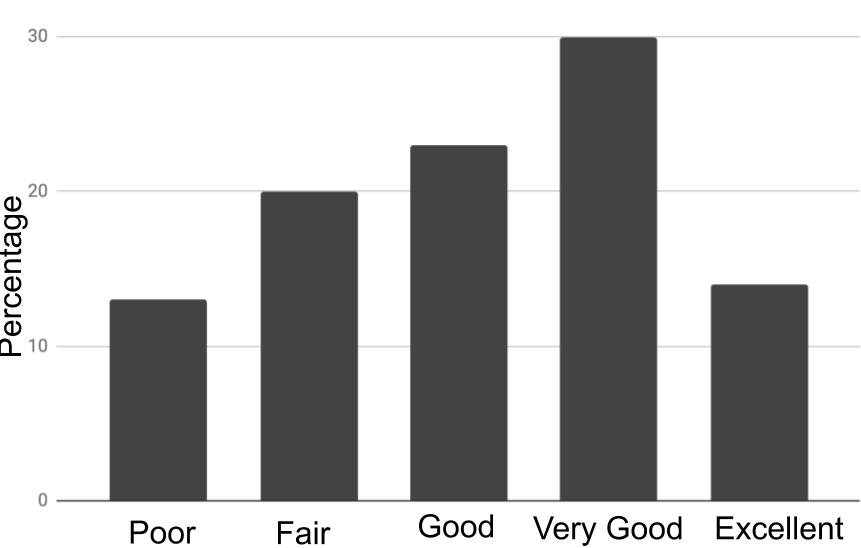}
  \caption{Super Turkers' view on the computed criteria. Most Super Turkers approved the criteria.}
  \label{fig:perspectiveSuperTurker}
  \vspace{-5mm}
\end{figure}

We also asked Super Turkers for feedback on the computed criteria (see Fig \ref{fig:perspectiveSuperTurker}). In general, Super Turkers approved the criteria. They also approved of doing HITs that paid slightly more than \$0.23 cents when no other information about the requester was available. Their logic, in such a situation, 
was to accept a task that paid slightly higher than the average HIT in order to start earning money rather than waiting. For instance, one Super Turker mentioned:

\emph{``Being somewhat informed about a requester and the quality of work will help you make more money. Sometimes you just have to go with it because otherwise you won't make anything.'' -Super Turker A}



Super Turkers commented that accepting HITs that offered lower than their preferred pay just to ensure income was at times necessary, especially as the market might not offer anything better:

\emph{``I think it's a good strategy, but sometimes you may have to do work for requesters whose hourly pay is lower. Sometimes you have to do what work is available.'' -Super Turker B}

The Super Turkers who found the criteria ``poor'' or ``fair'' were primarily workers whose strategy focused on doing batch HITs. They felt the criteria did not represent their process. This is expected given that the algorithm focused on computing a criteria that was most common; instead of something that was representative of all Super Turkers. But the criteria we identified served the purpose of being computationally appropriate in representing the majority.

\emph{``...Setting it at \$0.23 may cause workers to miss out on an excellent, one click batch that may be \$0.05...'' -Super Turker C}


Given these positive results, we tested the criteria with novices in the real world to see how it might help increase their wages.


\section{Novices Use Super Turker Criteria}
The first study allowed us to compute a common and concise criteria of how the average Super Turkers decided what transparency metrics to consider. 
The criteria was algorithmically constructed based on the input from Super Turkers. 

After computing the Super Turker criteria, we conducted a two-week field experiment on MTurk and investigated:
\begin{itemize}
\item Do novices following the Super Turker criteria perform individual tasks that pay more?
\item Do control group novices, who utilize their own criteria, discover upon the identical Super Turker criteria?
\item Do novices following the Super Turker criteria increase their hourly wage?
\end{itemize}

\subsection{Field Experiment: Methods}

Our field experiment followed a randomized control-group pretest-test design that is characterized by being similar to a between subjects study, but with the addition that measurements are taken both before and after a treatment \cite{dimitrov2003pretest}. This setup facilitates better understanding of the change generated from experiments. For this purpose, we split our experiment into two stages: a six-day pretest stage to understand novices' behavior and wages before our intervention, and a six-day test stage to understand novices' behavior and wages after our intervention (i.e., after telling novices to follow the criteria). We divided the subjects into two groups:
\begin{enumerate}
    \item \textbf{Control group.} During the entire study (i.e., throughout the pretest and test stages), novice crowd workers used transparency tools and completed tasks as normal; 
    \item \textbf{Experimental group.} In the pretest, novice crowd workers used transparency tools and completed tasks as normal; but, in the test stage, they were instructed to follow the decision-making criteria of Super Turkers.
\end{enumerate}

We recruited 100 novice Turkers. Similar to prior work, we considered novices were workers who had completed less than 500 HITs \cite{chiang2018crowd,suzuki2016atelier}. The completion of 500 HITs is also within the probation period of MTurk \cite{community_2017,pritam_2017} 
We randomized novices across each of our two conditions (50 workers in each groups). All novices in our study reported using Turkopticon and TurkerView (which is aligned with the findings of previous work that identified that most workers are using transparency tools \cite{kaplan2018striving}). 

{\bf Pretest Stage.} In this stage, participants across conditions were asked to: (1) install a Google Chrome plugin we developed; (2) perform HITs as normal. The plugin allowed us to track participants' behavior (types of tasks they did, requesters they worked with, and earnings made.) We used this information to study how workers' wages changed over time and track how they utilized transparency tools. Participants were allowed to uninstall our plugin at any time, and we rewarded them for the period of time that they had our plugin installed on their computer. We paid novices \$0.60 USD for installing our extension, accounting for the US federal minimum wage (\$7.25/hour) as installation took less than 4 minutes. 

{\bf Test Stage.} In this stage, novices in the control group were asked to continue working using their customary method; while in the experimental group, novices were asked to make decisions using the identified Super Turker criteria. Participants were also informed that the criteria came from Super Turkers and could help them to increase their hourly wages. For this purpose, at the beginning of the test stage, we emailed participants and informed them of the activity they would do with us that week. Similar to prior work that has run field experiments on MTurk, we paid participants in the experimental group another \$0.5 for reading the criteria and an additional bonus of \$0.10 each time they followed the criteria to select a HIT and completed it. On average, we paid participants a total of \$4.90 for following the criteria; the participant who received the most for following the criteria earned \$24.70. Similar to \cite{doroudi2016toward}, only participants in the experimental group were paid extra to follow the criteria. In both the control and experiment group, we continued paying participants each day they kept our plugin installed. To avoid our remuneration interfering with our study, we paid these bonuses at the end of our experiment and did not include our HIT remuneration when calculating workers' hourly wages.

\subsubsection{Collecting and Quantifying Workers' Behavior.}  
For our study, we needed methods for: (1) collecting and quantifying workers' behaviors and the HITs they performed; (2) flagging when workers utilized the Super Turker criteria; (3) measuring how much workers' hourly wages changed when following the criteria. We created a Google Chrome extension$^3$ (i.e. plugin) to collect crowd workers' behavior on Mturk. The plugin tracked the metadata about the HITs that workers previewed or accepted, and the timestamps of when workers accepted, submitted or returned HITs (i.e., tasks which were accepted but for various reasons not completed). In specific, the plugin collected:

\begin{itemize}
    \item HIT information, such as title, rewards, timestamps (accept/ submit/ return), requester IDs, HIT Group IDs, and HIT IDs.
    \item Worker information, such as daily earnings on the dashboard, installed extensions, approval rate, and worker IDs.
    \item Requester reputation information, such as ratings on Turkopticon and TurkerView. 
\end{itemize}

There were certain things our plugin did not record: the required specific qualifications for a HIT and the HITs that participants decided to not take (preview or accept). To maintain workers' privacy, our browser extension also did not record workers' browsing records outside of MTurk or workers' personal data (such as worker ID, qualifications, among other personal metrics).

\subsubsection{Identifying Whether Workers Follow The Criteria}

Once we had collected and quantified the completed tasks of novices, our goal was to identify the novices who had followed the Super Turker criteria. To accomplish this, we took the transparency data available for each task that novices completed (i.e., TurkerView's expected hourly wage, Turkopticon's fairness score, the HIT reward) and used the flowchart presented in section \ref{strategy} to determine whether the HIT followed the criteria. We name HITs that meet the established criteria as Super HITs. After this step, each novice had a list of Super HITs and non-Super HITs associated with them. We considered that novices with over 70\% of their HITs labeled as Super HITs to be workers who followed the Super Turker criteria. For our study, the threshold of 70\% was selected based on prior research that has established this amount as an adequate threshold to measure whether people are following new behavioral patterns \cite{lynch2015retrospective,Fogg:2009:BMP:1541948.1541999}. After this step, we had a list of novices in our experimental group who had followed (or not) the criteria. 


\begin{figure}
\centering
 \includegraphics[width=0.99\columnwidth]{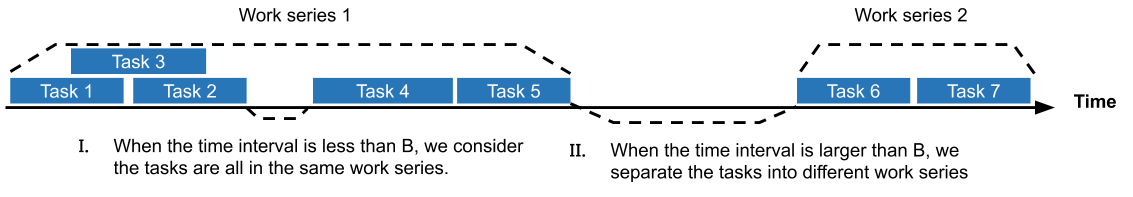}
 \vspace{-2mm}
 \caption{We separate work series based on time intervals: When the time interval is less than B, we consider the tasks are all within the same work series. If not, we separate them into different work series.}
 \label{fig:measure}
\end{figure}



\subsubsection{Measuring Workers' Hourly Wages} 
After compiling the list of novices who followed the criteria and those who did not, our goal was to calculate each novice's hourly wage and determine whether the workers following the criteria increased their wages. To calculate the hourly wages of a worker we need to measure: (a) the income that a worker earned; and (b) the amount of time they worked to earn that income. 

{\bf A. Total Earnings.} A worker's earnings does not come solely from HIT compensation (i.e., the salary that MTurk states they will pay when HITs are completed). Workers might also earn HIT bonuses, i.e., additional rewards from requesters whose amount is usually unknown before performing the HIT. To record both reward and bonuses adequately, we logged the ``$\text{Daily Income}$'' from workers' dashboard which already considers both values directly. Using the values specified in workers' dashboard helps us to consider circumstances where workers' labor might have been rejected, i.e., no payment for completing a HIT. However, there are some issues with tracking workers' wages using the dashboard: not all requesters pay workers as soon as they submit their work. In these cases, the payment can be delayed for several days. To overcome this problem, we checked workers' dashboard three days after the end of the whole experiment to give requesters time to make their corresponding payments. For all participants in our study we calculated their daily wages ($Income_D$) through this method for the twelve days they participated in our study. Using this method, we were able to calculate workers' total earnings. 

{\bf B. Total Working Time.} To calculate how much time a worker required to complete a task and earn specific wages, our extension logged:  $\text{Time}_{\text{start}}$: the exact moment when a worker accepts a HIT; and $\text{Time}_{\text{end}}$: the moment when a worker returns or submits a HIT. Notice that workers might take breaks \cite{labor2019break} and spend time searching for HITs, calculating the working time as simply $\text{Time}_{\text{end}}-\text{Time}_{\text{start}}$ is not appropriate. To overcome this problem, we adopt an approach similar to \cite{hara2018data}, where we consider that a worker is doing series of tasks continuously if their time interval is less than B minutes (as shown in Figure \ref{fig:measure}.I) and consider they have started a new series of tasks if the time interval is larger than B minutes (as shown in Figure \ref{fig:measure}.II). When calculating the hourly wage, we set the between time (B) between the task series as 12 minutes because in normal industries employees must be paid for any break \cite{labor2019break}. The tasks done within the same work series $S$ share the same time interval $\text{Time}_{S,d}$, where $d$ represents the date when that particular work series began. We measure the time interval for series $S$ as follows:
\begin{equation}
    \text{Time}_{S,d} = \max_{s\in{S}}\{\text{Time}_{\text{end},s,d}\} - \min_{s\in{S}}\{\text{Time}_{\text{start},s,d}\}.
\end{equation}

A worker's total number of work hours in day D, (${WorkingHour}_D$), is the sum of the working time of all the series they did on day D: 
\begin{equation}
    {WorkingHour}_D = \Sigma_{d\in D}^{} \text{Time}_{S,d}
\end{equation}

For a given worker $w$, her overall hourly wage for day $D$ is:
\begin{equation}
w_D = \frac{\text{Income}_D}{{WorkingHour}_D}.
\end{equation}

After this step, we had the hourly wages earned for all novices; and for novices in the experimental group, we labeled them according to whether or not they followed the criteria.

\subsection{Field Experiment: Findings}
Our experiment ran for 12 days during late October 2018. A total of 100 unique novice workers participated in our study and were randomized across our two conditions. Participants visited (i.e., previewed or accepted) a total of 25,899 HITs during the two week process. The recorded HITs belonged to 2,568 unique HIT groups posted by 1,394 unique requesters. Novices visited 261 HITs on average, and visited 102 HITs in the median.

\subsubsection{Novices, Super Turker Criteria, And Hourly Wages of HITs.}
In the previous section, we established a common and concise criteria constructed from surveying Super Turkers. However, it was unclear whether following the criteria would guide novices to perform individual HITs that actually paid them more per hour. 

Part of the criteria was to select tasks that TurkerView predicted would likely earn workers a particular hourly wage. However, it was yet unclear whether TurkerView offered available information on these individual HITs. 
To better understand the ecosystem in which our novices operated and contextualize the availability of transparency information, we analyzed the details of all the HITs novices completed. In our field experiment, our participants worked for 1,394 different requesters. Although, our statistics demonstrated that only 772 of these requesters had been reviewed on both TurkerView and Turkopticon, 500 requesters were not reviewed on TurkerView, 430 requesters were not on Turkopticon, and 318 requesters had not been reviewed on either TurkerView or Turkopticon. These results demonstrated that for 36\% of the requesters for whom novices worked with, there was no information about them on transparency tools. 
For requesters whose transparency information was missing, the Super Turker criteria recommends analyzing the reward of the HIT, i.e., the default metric on MTurk, and only performing HITs that paid more than \$0.23.


Additionally, even when TurkerView's expected hourly wage was available, it is calculated for average workers (i.e., TurkerView considers the average time it takes all workers to complete tasks for a particular requester and based on this calculates the expected hourly wage given what that requester typically pays.) However, it is unclear whether this average length of time would also pertain to novices. It could potentially take novices more time to complete certain tasks, and hence they would earn less.

In order to investigate whether novices following the Super Turker criteria performed tasks that actually paid them more per hour, we took all the HITs that novices completed and calculated the real hourly wage that novices received for the HITs. This was based on how much time it took them to perform the task, and the actual amount of money they received for finishing the task. Next, we identified the tasks that were Super HITs and those that were not, and we compared the hourly wage between these two groups. 

We first focused on inspecting the HITs that lacked transparency information. Across conditions, novices in our study accepted and submitted 9,503 HITs. Of these, there were 979 HITs from requesters who lacked data on Turkopticon and TurkerView. From this group there were 232 Super HITs (i.e., HITS that rewarded more than \$0.23) and 747 HITs that rewarded less than \$0.23 (non Super Hits). We conducted a Mann-Whitney U test to compare the real hourly pay between these two groups given that the sample sizes of these two kinds of HITs were different and they presented a large standard deviation. The Mann-Whitney U test showed that there was a significant difference in the median hourly wage between Super HITs and non-Super HITs when workers can only access the information about the rewards (U=68667, p<0.0001). The median hourly wage that Super HITs paid was \$3.97, while non-Super HITs paid \$2.76. Thus, following the SuperTurker criteria can guide novices in their decision-making to identify individual HITs that pay them higher hourly wages, even when transparency data about the requesters posting the HITs is missing.

\begin{figure}
\centering
   \includegraphics[width=0.99\columnwidth]{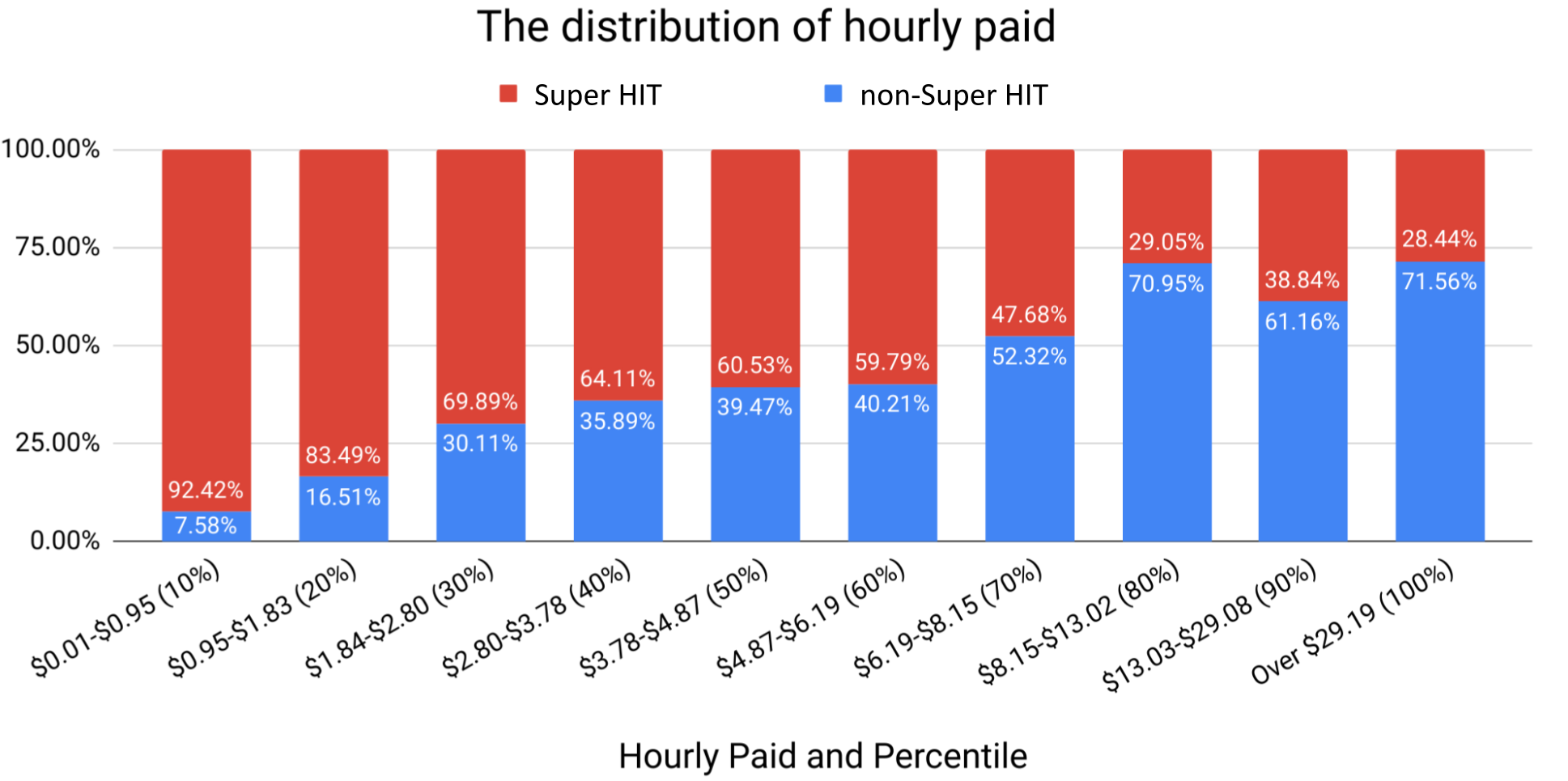}
   \vspace{-2mm}
   \caption{Distribution of the hourly wage of all HITs that novices selected that had transparency data available. Super HITs tended to provide a higher hourly wage.}
   \label{fig:criteria}
 \end{figure}
 
 
Next, we compared the real hourly wages of Super HITs and non-Super HITs that had available transparency data. Figure \ref{fig:criteria} showcases the hourly wage that workers earned for Super HITs (N=4,047, 42.6\% of the submitted tasks) and for the HITs that did not follow the criteria (N=5,456, 57.4\% of the submitted tasks). We eliminated the outliers and computed a t-test to compare the difference between these two groups. The t-test showed that there was a significant difference in the actual hourly wage that workers received for completing Super HITs and non-Super HITs (t(5806)=11.151, p < 0.0001). This highlights, that novices who follow the Super Turker criteria are able to distinguish which individual HITs, when performed, will pay them higher wages per hour.

\subsubsection{SEARCHING TIME AND THE SUPER TURKER CRITERIA} Our previous analysis uncovered that individual Super HITs had a higher hourly wage than non-Super HITs. In general, the median hourly wage of all Super HITs was \$7.04/h , while the median hourly wage of all non-Super HITs was \$3.27. 
Note this hourly wage only considers the time that workers spent in completing a task, not the uncompensated time that workers need to expend to search for HITs. Crowd markets have placed these costs, that were traditionally absorbed by companies, onto workers \cite{gray2019ghost}. Such costs include the time that workers spend searching for work. In a micro job atmosphere, these costs to workers pose a serious issue.

In this analysis, we were interested in studying whether the quest for Super HITs could possibly lead novices to spend a greater amount of time searching for work, hence reducing their hourly wages (considering that the hourly wage involves the time workers spend searching for work in addition to the time completing work.) In this setting, we considered that the searching time includes searching for HITs in the HIT pool, previewing HITs, and accepting HITs but not yet submitting them. We identified that novices for a given task spent a mean and median searching time of 168 seconds (SD=388) and 20 seconds, respectively. Additionally, the searching time for Super HITs and normal HITs was different. The median searching time for normal HITs was 9 seconds, while the median searching time for Super HITs was 61 seconds. Next we calculated novices' total working time as time spent searching for work plus time spent completing tasks. Through our analysis, we identified that the median hourly wage for Super HITs, when considering searching and working time together, was \$4.12/h; while for non-Super HITs this wage was \$2.13/h. Therefore, it appeared that while there was an overhead cost imposed upon workers for searching for Super HITs (i.e. avoiding non-Super HITs), the overhead cost was worth it for increasing the hourly wage. 



\subsubsection{Novices Discovering The Super Turker Criteria Independently}

Next, we examined whether novices had already adopted the decision-making criteria of the Super Turkers before instructing them to follow it. If this were the case, it might be unnecessary to teach novices how to effectively use transparency tools to decide which tasks might pay more. For this purpose, we analyzed the meta-data of all the HITs that novices decided to perform in the pretest stage (i.e., we studied the HITs novices accepted and submitted). From this, we inspected the number of HITs that were Super HITs. We identified that a minority of novice workers in our study (25\%) were already unwittingly utilizing the Super Turker criteria. 



\subsubsection{The Super Turker Criteria And Novices' Hourly Wages}

\begin{figure}
\centering
 \includegraphics[width=0.90\columnwidth]{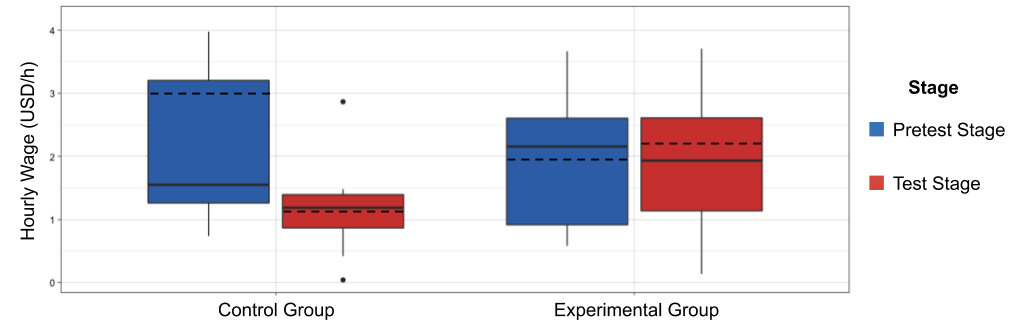}
 \vspace{-2mm}
 \caption{Box plot comparing the hourly wage of novices in both groups. The solid line denotes the median value and the dashed line denotes the mean value. 
 Novices following the Super Turker criteria increased their wages more.}
  \label{fig:boxplot}
\end{figure}


In this analysis, we focused on comparing the actual hourly wages that were received by novices following the Super Turker criteria compared to what control group novices received. However, 32 control group participants and 19 from the experimental group uninstalled our plugin before the test stage started. This was likely due to the lengthy nature of our study. For the remaining 49 participants, we identified that 7 control group participants were already using the Super Turker criteria from the beginning (i.e., during pretest stage), and 12 of the participants in the experimental group never followed the identified criteria. Given that we were interested in studying the change of wages that occurred after utilizing the Super Turker criteria, we discarded the above participants' data. In the end, 21 experimental group participants and 11 control group participants were remaining for our analysis. 

From these participants, we studied how much their hourly wage changed between the pretest and test stages. For all novices, we calculated the change of their hourly wage as the median hourly wage they received in the test stage minus their median hourly wage in the pretest stage. Fig \ref{fig:boxplot} presents an overview of how much the hourly wage of novices changed in both the control and the experimental group. During the pretest stage: the control group had a median hourly wage of \$1.55 and mean of \$3.00; while the experimental group had a median hourly wage of \$2.16 and mean of \$1.95.
In the test stage, the median and mean hourly wage that control group novices earned reduced to \$1.19 and \$1.17  respectively. Meanwhile, the median and mean hourly wage of novices in the experiment group increased slightly to \$1.93  and \$2.20  respectively. 

We computed a Mann-Whitney U test to examine whether the difference in observed wages between the control and experimental groups was significant. We used the Mann-Whitney U test given that both the control and experimental groups were independent, had different variances, and presented small sample sizes. Moreover, The Mann-Whitney U test does not compare mean scores but median scores of two samples. Thus, it is much more robust against outliers and heavy tail distributions. The Mann-Whitney U test showed that there were significant differences in how much the median hourly wages of novices varied between the control and experimental groups ($U=48, p=0.006$). 

Next, we investigated whether the change in wages that each condition presented between pretest and test stages was significant. A Wilcoxon signed-rank test highlighted that the change in wages in the experiment group was not significantly different (Z=115, p=1.00). However, the change in wages that the control group presented was significantly different (Z=61, p=0.04). This finding might hint that the general pool of HITs during the test stage had worse remuneration than the pretest; hence, this explains why we witnessed a decrease in control group wages since there was no intervention. To better understand this ecosystem, we examined the hourly wages of all HITs in the pretest and test stages. In general, the mean and median hourly pay of all HITs in the pretest stage was \$11.68/h (SD=44.92) and \$4.73/h. The mean and median hourly wage in the test stage was \$14.877 (SD=92.66) and \$4.59/h. Between stages, there was no significant difference in how much HITs paid novices per hour (U=5,280,800, p=0.34).

Given that the hourly wage of the HITs appeared to be in general the same in the pretest and test stages, it remained unclear why the control group had decreased their wages so drastically in the test stage. We decided to investigate further. We studied the amount of time workers spent searching for work in the pretest and the test stages. A change in searching time could denote that workers might have experienced a harder time finding tasks to perform, even if there was no change in the rewards that they received for the tasks. In general, workers could have seen a drop in their earnings simply due to not enough available tasks for that week. For this purpose, we calculated for each stage the average time that all workers spent searching for tasks. We discovered that there was a significant difference between the amount of time that workers spent searching for HITs in pretest vs the test stages  ($U=4,109,600, p<0.001$). Participants spent 158 seconds on average (median=16, SD=379) to find a HIT in the pretest stage; and 262 seconds (median=64, SD=436) to find a HIT in the test stage. This increment could be attributed to the fact that identifying Super HITs is more time consuming than searching for normal HITs. 

To understand more deeply what was taking place, we compared the difference in searching time for the Super HITs and non-Super HITs. 
In the pretest stage, the median searching time for Super HITs was 46 seconds; while it took 9 seconds for non-Super HITs. In the test stage, the median searching time that the control group took to identify Super HITs was 131 seconds; while it took 121 seconds for non-Super HITs. Note that we analyzed all HITs in the pretest stage; however, we only analyzed the control group HITs in the test stage. This was done to avoid the influence of our intervention (i.e., the experimental group in the test stages cannot represent the regular searching time due to the fact that we informed them to search according to the criteria). We identified that the searching time for both Super HITs and non-Super HITs increased from pretest stage to test stage. This result suggests that, in general, MTurk likely had fewer HITs available for novices in the test stage. This was why, on average, the novices searching time increased. However, despite this adversity, novices following the Super Turker criteria maintained their wage level, while the wages in the control group decreased significantly.

\section{Discussion}

Our experiments demonstrate the potential of using the criteria of Super Turkers to guide novices on how to use transparency tools to find work on MTurk. The majority of novices following the Super Turker criteria increased their wages, even while novices in the control condition decreased their salaries (likely due to the limited tasks available that week on MTurk.) Our study provides insights into the impact these highly effective workers can have on novices, as well as demonstrating the feasibility of connecting transparency tools with educational opportunities to increase workers' wages. In this section we discuss our results, highlighting especially the challenges and opportunities we envision in the research area.

\subsubsection*{\bf Super Turker Criteria.} Our study uncovered one of the most common and concise set of criteria that Super Turkers adopted to handle transparency to find fair work. It was computationally derived from surveys of Super Turkers and helped novices decide which MTurk tasks to perform by utilizing transparency tools as a means to earn higher wages. Our model was based on previous studies that demonstrated how ``shepherding'' novices could help them to improve their labor \cite{dow2011shepherding}. 


An interesting observation on the particular Super Turker criteria that we uncovered, is that it considered the circumstance that when the hourly wage metric was missing, it was best to look at Turkopticon's fairness value instead of inspecting other metrics related to how well the requester paid. Fairness on Turkopticon regards whether the requester will reject or accept a worker's labor. Crowd markets have traditionally contained power imbalances where workers have limited power in comparison with requesters and platform owners \cite{gray2019ghost}. Part of the imbalances arise because most crowd markets are very concentrated: almost 99\% of all tasks are posted by 10\% of the requesters, who do not have to negotiate with workers about whether or not they will accept or reject their labor. 
If workers and their work are rejected by a requester, workers generally suffer from a lack of accountability in response to their complaints or attempts at restitution from requesters and platform owners. 
This can translate into wasted time, loss of a paycheck, and little opportunity to raise awareness about possible exploitation \cite{silberman2018responsible}. According to Pew research, 30\% of all gig workers have experienced non-payment at least once \cite{gray2019ghost}. The US Freelancer Union reports a much larger number, where allegedly 70\% of current freelancers have at least one client who has not paid them \cite{costs_2019}. This could explain why the criteria recommended inspecting Turkopticon's fairness score. This step in the criteria, in particular, attempted to ascertain which situations needed to be avoided by Turkers in order to prevent non-payment or rejected labor.

The Super Turker criteria we identified considered that if the Turkopticon fairness score and the TurkerView hourly wage were missing, the best course of action was to look directly at the HIT's reward. One reason for this strategic behavior by Super Turkers is that for every minute they spend analyzing a HIT's transparency metrics, they are losing out on the chance to earn money (ultimately reducing their hourly wages). Notice also that lower reward HITs could still provide high hourly wages, especially because the distribution of the hourly wage is a near normal distribution  given a specific reward amount \cite{hara2018data}. In this setting, Super Turkers might be betting that HITs that are paying more than \$0.23 will likely provide higher hourly wages. Moreover, following this behavior will likely reduce the unpaid time spent searching for labor.


\subsubsection*{\bf Field Experiment.} 
While transparency is now available to Turkers via different plugins, workers still receive less than the minimum wage \cite{hara2018data,stiglitz2000contributions}. We believe that part of the problem is that workers likely cannot interpret or evaluate the value of their work and its relationship with the transparent information they are now observing. Several workers might also not necessarily have the analytical skills to interpret all of the transparency metrics that tools, like Turker View,  provide \cite{jarrahi2019algorithmic}.  Our aim with this research was to identify a practical way to use transparency in MTurk and study if that could help crowd workers to increase their salaries. Our field experiment highlighted that guiding novices in their decision-making by following the Super Turker criteria lead them to earn wages that were higher than what they would earn working on their own. Our results emphasize that transparency is required but it is not sufficient. Utilizing transparency skillfully can transform salaries on MTurk. Our method helped novices to earn more. We identified that the difference in hourly wage between the control group and the experimental group was significant. 

Observing the decrease in wages of the control group during the test stage, we recognize that workers' salaries are dependent on task availability. Thus, we suspect the hourly wage decline may be attributed to the fluctuation of the task pool. The composition of the tasks available to workers was different each day during the time period we ran our experiment. The fact that tasks are different from day to day in crowd work is documented \cite{kaplan2018striving}. MTurk does not guarantee that there are enough well-paid tasks every day and this issue poses a difficulty for crowd workers\cite{berg2018digital}. Through our analysis we identified that novices in the control group spent a greater amount of time searching for HITs during the test stage. The time spent searching for work is time where workers are not paid. Hence, novices in the control condition saw their hourly wages reduced as they spend more unpaid time searching for work. However, even during a phase where there might have been limited high paying tasks, our experimental group was able to increase their wages. 

As we think in practice about how to deploy these strategies at scale \cite{krafft2019defining}, it can be important to consider how enforcing strategies that make use of commercial tools, such as Turker View, could potentially create further social divisions on MTurk \cite{williams2019perpetual} (especially, as only workers who could afford to pay for such tools could follow the strategy). Moving forward, we plan to explore strategies with only opensource platforms and that are aware of workers' privacy and autonomy concerns\cite{gray2019ghost,jarrahi2019platformic,lopez2017behind}. 

\subsubsection*{\bf Difficulties in promoting the Super Turker Criteria.} Our field experiment also helped us to identify first hand the difficulties in getting novices to follow the decision-making criteria from Super Turkers (even though following the criteria had the added incentive of potentially higher wages). Half of the novices in our experimental group did not follow the instructions from Super Turkers. One of the reasons for this might be that novices simply did not have the qualifications to perform the HITs that satisfied the criteria that Super Turkers recommended. Therefore, even if they wanted to follow the criteria they might not have been able to access and do the related HITs. In order to protect workers' privacy, our plugin did not track worker qualifications or the qualifications that tasks imposed on workers, we believe there is likely value in exploring interfaces that help crowd workers gain the qualifications they need to access higher paying jobs \cite{kaplan2018striving}. However, it is interesting to note that 75\% of the Super Turkers in our study lacked master Turker qualifications. It is unclear which type of qualifications novices should strive for in order to successfully perform higher paying tasks. Future work could explore how qualifications affect the type of labor and wages novices can access. 

Another reason why workers might not have followed the Super Turker criteria is that we posted and shared the criteria as requesters. Workers might have viewed the Super Turker instructions as part of the task they were doing for us and not as something that was meant to really support them. As a result, their motivation for following the criteria might have been lacking. Previous education research has shown that students who use reciprocal teaching strategies (i.e., strategies where students share with each other advice) tend to have better performance than students working on their own or even students who are working directly with a professor  \cite{shadiev2014effects}. Similarly, recent crowdsourcing research has highlighted that increasing the interactions between workers can help novices to more easily develop their skills and grow \cite{suzuki2016atelier,chiang2018crowd,dontcheva2014combining}. In future work, we will examine different interfaces for sharing Super Turker criteria with workers to increase the adoption of the criteria. 

\subsubsection*{\bf Design Implications and Future Work.} We observed that novices could improve their wages by imitating how experienced workers used transparency tools. We believe there is value for researchers and practitioners to build systems that teach novices the transparent information that is pertinent to achieving their goals. The objectives of these systems are not only to recommend to novices the tasks they should perform, but also to help them understand and learn what kind of tasks pay fairly (potentially extrapolating such knowledge to other crowd markets and online spaces). We also believe there is value in creating on-the-job tutorials that can guide workers on the type of labor they should perform. 
Designing such tutorials is time-consuming but crucial for empowering workers, who often have limited time and resources, to earn higher wages. Future work could explore designing data driven tutorials that are generated in part based on the patterns of effective workers. There might also be value in exploring educational material that has been generated for audiences with time constraints \cite{escobedo2012mosoco}. 

Additionally, to build more trust and participation on crowd markets, it might also be worth to explore transparent interfaces that can inform the different actor of a marketplace just how much each actor is being fair and respectful of others' values \cite{chiang2018exploring}. Future work could explore other ways of recruiting Super Turkers and eliciting information from them, e.g., via video recordings or interviews\cite{park2014toward}. Such studies could explore how using different mechanisms for eliciting information relates to the type of information that is obtained from Super Turkers. Related, it might be interesting to specifically investigate other types of Super Turker criteria that might exist. For instance, investigate how Super Turkers use transparency tools when multitasking\cite{williams2019perpetual}. Future work could also explore what happens in other crowd platforms (e.g., Uber, Upwork, or Citizen Science platforms) when novices adopt the strategies from high earning participants or the strategies from accounts who are contributing the most \cite{crowston2019coordinating,luthersolving,lampinen2016hosting,radford2016volunteer,savage2016botivist}. 

{\bf Limitations.}
We conducted a real world experiment which is not simple given the limited availability of HITs and lack of information provided by MTurk about workers' hourly wages \cite{hara2018data}. The issue of limited and inconsistent HIT availability has been documented in other research and we experienced, firsthand, the possible implications this has on workers' wages and on conducting ``clean'' research \cite{berg2018digital}. 
Notice also that we recruited Super Turkers who were willing to engage in surveys on MTurk (missing those who do not do surveys). Additionally, our algorithm focused on computing a criteria that was commonly used and simple to implement. 
Therefore, our criteria did not represent all Super Turker behavior. For instance, some Super Turkers might only do HIT batches that pay \$0.01 cent and can be completed in less than 10 seconds, resulting in an hourly wage of approximately \$36/hour. 
Nonetheless, given that our goal was to identify one of the strategies that Super Turkers adopted, and study how it plays out when used by novices, we considered our approach to be computationally appropriate and representative. 
Notice that our study focused on breadth instead of depth to start to shed needed light on how Super Turkers use transparency and how this plays out when adopted by novices. Future work could conduct longitudinal studies inspecting the amount of time it takes novices to adopt on their own some of the Super Turker strategies vs guiding them to adopt the strategies from the start.  Having task interruptions and multitasking are dependent on people's work style and preferences \cite{lascau2019monotasking,williams2019perpetual}. Future work could explore how strategies with multitasking differ from strategies with only monotasking.

{\bf Acknowledgements.} {Special thanks to Amy Ruckes for the immense
feedback and iterations on this work. Thanks to Caroline Anderson, Pankaj Ajit for helping us to start exploring this area. This work was partially supported by NSF grant FW-HTF-19541.}
\bibliographystyle{ACM-Reference-Format}
\bibliography{acmart}


\begin{thebibliography}{81}


\ifx \showCODEN    \undefined \def \showCODEN     #1{\unskip}     \fi
\ifx \showDOI      \undefined \def \showDOI       #1{#1}\fi
\ifx \showISBNx    \undefined \def \showISBNx     #1{\unskip}     \fi
\ifx \showISBNxiii \undefined \def \showISBNxiii  #1{\unskip}     \fi
\ifx \showISSN     \undefined \def \showISSN      #1{\unskip}     \fi
\ifx \showLCCN     \undefined \def \showLCCN      #1{\unskip}     \fi
\ifx \shownote     \undefined \def \shownote      #1{#1}          \fi
\ifx \showarticletitle \undefined \def \showarticletitle #1{#1}   \fi
\ifx \showURL      \undefined \def \showURL       {\relax}        \fi
\providecommand\bibfield[2]{#2}
\providecommand\bibinfo[2]{#2}
\providecommand\natexlab[1]{#1}
\providecommand\showeprint[2][]{arXiv:#2}

\bibitem[\protect\citeauthoryear{??}{com}{2017}]%
        {community_2017}
 \bibinfo{year}{2017}\natexlab{}.
\newblock \bibinfo{title}{[Guide] - Welcome to the world of Mechanical Turk}.
\newblock
\newblock
\urldef\tempurl%
\url{https://forum.turkerview.com/threads/welcome-to-the-world-of-mechanical-turk.112/}
\showURL{%
Retrieved June 26, 2019 from \tempurl}


\bibitem[\protect\citeauthoryear{aaai Fall Symosium~2019}{aaai Fall
  Symosium~2019}{[n.d.]}]%
        {luthersolving}
\bibfield{author}{\bibinfo{person}{aaai Fall Symosium~2019}.}
  \bibinfo{year}{[n.d.]}\natexlab{}.
\newblock \showarticletitle{Solving AI's last-mile problem with crowd-augmented
  expert work}.
\newblock  (\bibinfo{year}{[n.\,d.]}).
\newblock


\bibitem[\protect\citeauthoryear{Berg}{Berg}{2015}]%
        {berg2015income}
\bibfield{author}{\bibinfo{person}{Janine Berg}.}
  \bibinfo{year}{2015}\natexlab{}.
\newblock \showarticletitle{Income security in the on-demand economy: Findings
  and policy lessons from a survey of crowdworkers}.
\newblock \bibinfo{journal}{\emph{Comp. Lab. L. \& Pol'y J.}}
  \bibinfo{volume}{37} (\bibinfo{year}{2015}), \bibinfo{pages}{543}.
\newblock


\bibitem[\protect\citeauthoryear{Berg, Furrer, Harmon, Rani, and
  Silberman}{Berg et~al\mbox{.}}{2018}]%
        {berg2018digital}
\bibfield{author}{\bibinfo{person}{Janine Berg}, \bibinfo{person}{Marianne
  Furrer}, \bibinfo{person}{Ellie Harmon}, \bibinfo{person}{Uma Rani}, {and}
  \bibinfo{person}{M~Six Silberman}.} \bibinfo{year}{2018}\natexlab{}.
\newblock \showarticletitle{Digital labour platforms and the future of work:
  Towards decent work in the online world}.
\newblock \bibinfo{journal}{\emph{Geneva: International Labour Organization}}
  (\bibinfo{year}{2018}).
\newblock


\bibitem[\protect\citeauthoryear{Bergvall-K{\aa}reborn and
  Howcroft}{Bergvall-K{\aa}reborn and Howcroft}{2014}]%
        {bergvall2014mazon}
\bibfield{author}{\bibinfo{person}{Birgitta Bergvall-K{\aa}reborn} {and}
  \bibinfo{person}{Debra Howcroft}.} \bibinfo{year}{2014}\natexlab{}.
\newblock \showarticletitle{Amazon Mechanical Turk and the commodification of
  labour}.
\newblock \bibinfo{journal}{\emph{New Technology, Work and Employment}}
  \bibinfo{volume}{29}, \bibinfo{number}{3} (\bibinfo{year}{2014}),
  \bibinfo{pages}{213--223}.
\newblock


\bibitem[\protect\citeauthoryear{Bigham, Jayant, Ji, Little, Miller, Miller,
  Miller, Tatarowicz, White, White, et~al\mbox{.}}{Bigham
  et~al\mbox{.}}{2010}]%
        {bigham2010vizwiz}
\bibfield{author}{\bibinfo{person}{Jeffrey~P Bigham},
  \bibinfo{person}{Chandrika Jayant}, \bibinfo{person}{Hanjie Ji},
  \bibinfo{person}{Greg Little}, \bibinfo{person}{Andrew Miller},
  \bibinfo{person}{Robert~C Miller}, \bibinfo{person}{Robin Miller},
  \bibinfo{person}{Aubrey Tatarowicz}, \bibinfo{person}{Brandyn White},
  \bibinfo{person}{Samual White}, {et~al\mbox{.}}}
  \bibinfo{year}{2010}\natexlab{}.
\newblock \showarticletitle{VizWiz: nearly real-time answers to visual
  questions}. In \bibinfo{booktitle}{\emph{Proceedings of the 23nd annual ACM
  symposium on User interface software and technology}}. ACM,
  \bibinfo{pages}{333--342}.
\newblock


\bibitem[\protect\citeauthoryear{Bohanec}{Bohanec}{2009}]%
        {bohanec2009decision}
\bibfield{author}{\bibinfo{person}{Marko Bohanec}.}
  \bibinfo{year}{2009}\natexlab{}.
\newblock \showarticletitle{Decision making: A computer-science and
  information-technology viewpoint}.
\newblock \bibinfo{journal}{\emph{Interdisciplinary Description of Complex
  Systems: INDECS}} \bibinfo{volume}{7}, \bibinfo{number}{2}
  (\bibinfo{year}{2009}), \bibinfo{pages}{22--37}.
\newblock


\bibitem[\protect\citeauthoryear{Bohannon}{Bohannon}{2011}]%
        {bohannon2011social}
\bibfield{author}{\bibinfo{person}{John Bohannon}.}
  \bibinfo{year}{2011}\natexlab{}.
\newblock \bibinfo{title}{Social science for pennies}.
\newblock
\newblock


\bibitem[\protect\citeauthoryear{Callison-Burch}{Callison-Burch}{2014}]%
        {callison2014crowd}
\bibfield{author}{\bibinfo{person}{Chris Callison-Burch}.}
  \bibinfo{year}{2014}\natexlab{}.
\newblock \showarticletitle{Crowd-workers: Aggregating information across
  turkers to help them find higher paying work}. In
  \bibinfo{booktitle}{\emph{Second AAAI Conference on Human Computation and
  Crowdsourcing}}.
\newblock


\bibitem[\protect\citeauthoryear{Casler, Bickel, and Hackett}{Casler
  et~al\mbox{.}}{2013}]%
        {casler2013separate}
\bibfield{author}{\bibinfo{person}{Krista Casler}, \bibinfo{person}{Lydia
  Bickel}, {and} \bibinfo{person}{Elizabeth Hackett}.}
  \bibinfo{year}{2013}\natexlab{}.
\newblock \showarticletitle{Separate but equal? A comparison of participants
  and data gathered via Amazon’s MTurk, social media, and face-to-face
  behavioral testing}.
\newblock \bibinfo{journal}{\emph{Computers in human behavior}}
  \bibinfo{volume}{29}, \bibinfo{number}{6} (\bibinfo{year}{2013}),
  \bibinfo{pages}{2156--2160}.
\newblock


\bibitem[\protect\citeauthoryear{Chiang, Betanzos, and Savage}{Chiang
  et~al\mbox{.}}{2018a}]%
        {chiang2018exploring}
\bibfield{author}{\bibinfo{person}{Chun-Wei Chiang}, \bibinfo{person}{Eber
  Betanzos}, {and} \bibinfo{person}{Saiph Savage}.}
  \bibinfo{year}{2018}\natexlab{a}.
\newblock \showarticletitle{Exploring blockchain for trustful collaborations
  between immigrants and governments}. In \bibinfo{booktitle}{\emph{Extended
  Abstracts of the 2018 CHI Conference on Human Factors in Computing Systems}}.
  \bibinfo{pages}{1--6}.
\newblock


\bibitem[\protect\citeauthoryear{Chiang, Kasunic, and Savage}{Chiang
  et~al\mbox{.}}{2018b}]%
        {chiang2018crowd}
\bibfield{author}{\bibinfo{person}{Chun-Wei Chiang}, \bibinfo{person}{Anna
  Kasunic}, {and} \bibinfo{person}{Saiph Savage}.}
  \bibinfo{year}{2018}\natexlab{b}.
\newblock \showarticletitle{Crowd Coach: Peer Coaching for Crowd Workers' Skill
  Growth}.
\newblock \bibinfo{journal}{\emph{Proceedings of the ACM on Human-Computer
  Interaction}} \bibinfo{volume}{2}, \bibinfo{number}{CSCW}
  (\bibinfo{year}{2018}), \bibinfo{pages}{37}.
\newblock


\bibitem[\protect\citeauthoryear{ChrisTurk}{ChrisTurk}{2018}]%
        {turkerview}
\bibfield{author}{\bibinfo{person}{ChrisTurk}.}
  \bibinfo{year}{2018}\natexlab{}.
\newblock \bibinfo{booktitle}{\emph{TurkerView}}.
\newblock
\urldef\tempurl%
\url{https://turkerview.com/}
\showURL{%
Retrieved November 4, 2018 from \tempurl}


\bibitem[\protect\citeauthoryear{Crowston, Mitchell, and {\O}sterlund}{Crowston
  et~al\mbox{.}}{2019}]%
        {crowston2019coordinating}
\bibfield{author}{\bibinfo{person}{Kevin Crowston}, \bibinfo{person}{Erica
  Mitchell}, {and} \bibinfo{person}{Carsten {\O}sterlund}.}
  \bibinfo{year}{2019}\natexlab{}.
\newblock \showarticletitle{Coordinating advanced crowd work: Extending citizen
  science}.
\newblock \bibinfo{journal}{\emph{Citizen Science: Theory and Practice}}
  \bibinfo{volume}{4}, \bibinfo{number}{1} (\bibinfo{year}{2019}).
\newblock


\bibitem[\protect\citeauthoryear{Cushing}{Cushing}{2012}]%
        {cushing2012dawn}
\bibfield{author}{\bibinfo{person}{Ellen Cushing}.}
  \bibinfo{year}{2012}\natexlab{}.
\newblock \showarticletitle{Dawn of the digital sweatshop}.
\newblock \bibinfo{journal}{\emph{East Bay Express}}  \bibinfo{volume}{1}
  (\bibinfo{year}{2012}).
\newblock


\bibitem[\protect\citeauthoryear{De~Stefano}{De~Stefano}{2015}]%
        {de2015rise}
\bibfield{author}{\bibinfo{person}{Valerio De~Stefano}.}
  \bibinfo{year}{2015}\natexlab{}.
\newblock \showarticletitle{The rise of the just-in-time workforce: On-demand
  work, crowdwork, and labor protection in the gig-economy}.
\newblock \bibinfo{journal}{\emph{Comp. Lab. L. \& Pol'y J.}}
  \bibinfo{volume}{37} (\bibinfo{year}{2015}), \bibinfo{pages}{471}.
\newblock


\bibitem[\protect\citeauthoryear{Deng, Dong, Socher, Li, Li, and Fei-Fei}{Deng
  et~al\mbox{.}}{2009}]%
        {deng2009imagenet}
\bibfield{author}{\bibinfo{person}{Jia Deng}, \bibinfo{person}{Wei Dong},
  \bibinfo{person}{Richard Socher}, \bibinfo{person}{Li-Jia Li},
  \bibinfo{person}{Kai Li}, {and} \bibinfo{person}{Li Fei-Fei}.}
  \bibinfo{year}{2009}\natexlab{}.
\newblock \showarticletitle{Imagenet: A large-scale hierarchical image
  database}. In \bibinfo{booktitle}{\emph{2009 IEEE conference on computer
  vision and pattern recognition}}. Ieee, \bibinfo{pages}{248--255}.
\newblock


\bibitem[\protect\citeauthoryear{Dimitrov and Rumrill~Jr}{Dimitrov and
  Rumrill~Jr}{2003}]%
        {dimitrov2003pretest}
\bibfield{author}{\bibinfo{person}{Dimiter~M Dimitrov} {and}
  \bibinfo{person}{Phillip~D Rumrill~Jr}.} \bibinfo{year}{2003}\natexlab{}.
\newblock \showarticletitle{Pretest-posttest designs and measurement of
  change}.
\newblock \bibinfo{journal}{\emph{Work}} \bibinfo{volume}{20},
  \bibinfo{number}{2} (\bibinfo{year}{2003}), \bibinfo{pages}{159--165}.
\newblock


\bibitem[\protect\citeauthoryear{Dontcheva, Morris, Brandt, and
  Gerber}{Dontcheva et~al\mbox{.}}{2014}]%
        {dontcheva2014combining}
\bibfield{author}{\bibinfo{person}{Mira Dontcheva}, \bibinfo{person}{Robert~R
  Morris}, \bibinfo{person}{Joel~R Brandt}, {and} \bibinfo{person}{Elizabeth~M
  Gerber}.} \bibinfo{year}{2014}\natexlab{}.
\newblock \showarticletitle{Combining crowdsourcing and learning to improve
  engagement and performance}. In \bibinfo{booktitle}{\emph{Proceedings of the
  32nd Annual ACM Conference on Human Factors in Computing Systems}}. ACM,
  \bibinfo{pages}{3379--3388}.
\newblock


\bibitem[\protect\citeauthoryear{Doroudi, Kamar, Brunskill, and
  Horvitz}{Doroudi et~al\mbox{.}}{2016}]%
        {doroudi2016toward}
\bibfield{author}{\bibinfo{person}{Shayan Doroudi}, \bibinfo{person}{Ece
  Kamar}, \bibinfo{person}{Emma Brunskill}, {and} \bibinfo{person}{Eric
  Horvitz}.} \bibinfo{year}{2016}\natexlab{}.
\newblock \showarticletitle{Toward a learning science for complex crowdsourcing
  tasks}. In \bibinfo{booktitle}{\emph{Proceedings of the 2016 CHI Conference
  on Human Factors in Computing Systems}}. ACM, \bibinfo{pages}{2623--2634}.
\newblock


\bibitem[\protect\citeauthoryear{Dow, Kulkarni, Bunge, Nguyen, Klemmer, and
  Hartmann}{Dow et~al\mbox{.}}{2011}]%
        {dow2011shepherding}
\bibfield{author}{\bibinfo{person}{Steven Dow}, \bibinfo{person}{Anand
  Kulkarni}, \bibinfo{person}{Brie Bunge}, \bibinfo{person}{Truc Nguyen},
  \bibinfo{person}{Scott Klemmer}, {and} \bibinfo{person}{Bj{\"o}rn Hartmann}.}
  \bibinfo{year}{2011}\natexlab{}.
\newblock \showarticletitle{Shepherding the crowd: managing and providing
  feedback to crowd workers}. In \bibinfo{booktitle}{\emph{CHI'11 Extended
  Abstracts on Human Factors in Computing Systems}}. ACM,
  \bibinfo{pages}{1669--1674}.
\newblock


\bibitem[\protect\citeauthoryear{Escobedo, Nguyen, Boyd, Hirano, Rangel,
  Garcia-Rosas, Tentori, and Hayes}{Escobedo et~al\mbox{.}}{2012}]%
        {escobedo2012mosoco}
\bibfield{author}{\bibinfo{person}{Lizbeth Escobedo}, \bibinfo{person}{David~H
  Nguyen}, \bibinfo{person}{LouAnne Boyd}, \bibinfo{person}{Sen Hirano},
  \bibinfo{person}{Alejandro Rangel}, \bibinfo{person}{Daniel Garcia-Rosas},
  \bibinfo{person}{Monica Tentori}, {and} \bibinfo{person}{Gillian Hayes}.}
  \bibinfo{year}{2012}\natexlab{}.
\newblock \showarticletitle{MOSOCO: a mobile assistive tool to support children
  with autism practicing social skills in real-life situations}. In
  \bibinfo{booktitle}{\emph{Proceedings of the SIGCHI Conference on Human
  Factors in Computing Systems}}. \bibinfo{pages}{2589--2598}.
\newblock


\bibitem[\protect\citeauthoryear{Fogg}{Fogg}{2009}]%
        {Fogg:2009:BMP:1541948.1541999}
\bibfield{author}{\bibinfo{person}{BJ Fogg}.} \bibinfo{year}{2009}\natexlab{}.
\newblock \showarticletitle{A Behavior Model for Persuasive Design}. In
  \bibinfo{booktitle}{\emph{Proceedings of the 4th International Conference on
  Persuasive Technology}} \emph{(\bibinfo{series}{Persuasive '09})}.
  \bibinfo{publisher}{ACM}, \bibinfo{address}{New York, NY, USA}, Article
  \bibinfo{articleno}{40}, \bibinfo{numpages}{7}~pages.
\newblock
\showISBNx{978-1-60558-376-1}
\urldef\tempurl%
\url{https://doi.org/10.1145/1541948.1541999}
\showDOI{\tempurl}


\bibitem[\protect\citeauthoryear{Forman}{Forman}{1990}]%
        {forman1990multi}
\bibfield{author}{\bibinfo{person}{Ernest~H Forman}.}
  \bibinfo{year}{1990}\natexlab{}.
\newblock \showarticletitle{Multi criteria decision making and the analytic
  hierarchy process}.
\newblock In \bibinfo{booktitle}{\emph{Readings in multiple criteria decision
  aid}}. \bibinfo{publisher}{Springer}, \bibinfo{pages}{295--318}.
\newblock


\bibitem[\protect\citeauthoryear{Gadiraju, Checco, Gupta, and
  Demartini}{Gadiraju et~al\mbox{.}}{2017}]%
        {gadiraju2017modus}
\bibfield{author}{\bibinfo{person}{Ujwal Gadiraju}, \bibinfo{person}{Alessandro
  Checco}, \bibinfo{person}{Neha Gupta}, {and} \bibinfo{person}{Gianluca
  Demartini}.} \bibinfo{year}{2017}\natexlab{}.
\newblock \showarticletitle{Modus operandi of crowd workers: The invisible role
  of microtask work environments}.
\newblock \bibinfo{journal}{\emph{Proceedings of the ACM on Interactive,
  Mobile, Wearable and Ubiquitous Technologies}} \bibinfo{volume}{1},
  \bibinfo{number}{3} (\bibinfo{year}{2017}), \bibinfo{pages}{1--29}.
\newblock


\bibitem[\protect\citeauthoryear{Gadiraju and Demartini}{Gadiraju and
  Demartini}{2019}]%
        {gadiraju2019understanding}
\bibfield{author}{\bibinfo{person}{Ujwal Gadiraju} {and}
  \bibinfo{person}{Gianluca Demartini}.} \bibinfo{year}{2019}\natexlab{}.
\newblock \showarticletitle{Understanding Worker Moods and Reactions to
  Rejection in Crowdsourcing}. In \bibinfo{booktitle}{\emph{Proceedings of the
  30th ACM Conference on Hypertext and Social Media}}.
  \bibinfo{pages}{211--220}.
\newblock


\bibitem[\protect\citeauthoryear{Gadiraju, Fetahu, and Kawase}{Gadiraju
  et~al\mbox{.}}{2015}]%
        {gadiraju2015training}
\bibfield{author}{\bibinfo{person}{Ujwal Gadiraju}, \bibinfo{person}{Besnik
  Fetahu}, {and} \bibinfo{person}{Ricardo Kawase}.}
  \bibinfo{year}{2015}\natexlab{}.
\newblock \showarticletitle{Training workers for improving performance in
  crowdsourcing microtasks}.
\newblock In \bibinfo{booktitle}{\emph{Design for Teaching and Learning in a
  Networked World}}. \bibinfo{publisher}{Springer}, \bibinfo{pages}{100--114}.
\newblock


\bibitem[\protect\citeauthoryear{Gray and Suri}{Gray and Suri}{2019}]%
        {gray2019ghost}
\bibfield{author}{\bibinfo{person}{Mary Gray} {and} \bibinfo{person}{Siddharth
  Suri}.} \bibinfo{year}{2019}\natexlab{}.
\newblock \bibinfo{title}{Ghost Work: How to Stop Silicon Valley from Building
  a New Global Underclass}.
\newblock
\newblock


\bibitem[\protect\citeauthoryear{Han, Roitero, Gadiraju, Sarasua, Checco,
  Maddalena, and Demartini}{Han et~al\mbox{.}}{2019}]%
        {han2019all}
\bibfield{author}{\bibinfo{person}{Lei Han}, \bibinfo{person}{Kevin Roitero},
  \bibinfo{person}{Ujwal Gadiraju}, \bibinfo{person}{Cristina Sarasua},
  \bibinfo{person}{Alessandro Checco}, \bibinfo{person}{Eddy Maddalena}, {and}
  \bibinfo{person}{Gianluca Demartini}.} \bibinfo{year}{2019}\natexlab{}.
\newblock \showarticletitle{All Those Wasted Hours: On Task Abandonment in
  Crowdsourcing}. In \bibinfo{booktitle}{\emph{Proceedings of the Twelfth ACM
  International Conference on Web Search and Data Mining}}. ACM,
  \bibinfo{pages}{321--329}.
\newblock


\bibitem[\protect\citeauthoryear{Hara, Adams, Milland, Savage, Callison-Burch,
  and Bigham}{Hara et~al\mbox{.}}{2018}]%
        {hara2018data}
\bibfield{author}{\bibinfo{person}{Kotaro Hara}, \bibinfo{person}{Abigail
  Adams}, \bibinfo{person}{Kristy Milland}, \bibinfo{person}{Saiph Savage},
  \bibinfo{person}{Chris Callison-Burch}, {and} \bibinfo{person}{Jeffrey~P
  Bigham}.} \bibinfo{year}{2018}\natexlab{}.
\newblock \showarticletitle{A Data-Driven Analysis of Workers' Earnings on
  Amazon Mechanical Turk}. In \bibinfo{booktitle}{\emph{Proceedings of the 2018
  CHI Conference on Human Factors in Computing Systems}}. ACM,
  \bibinfo{pages}{449}.
\newblock


\bibitem[\protect\citeauthoryear{Harmon and Silberman}{Harmon and
  Silberman}{2018}]%
        {harmon2018rating}
\bibfield{author}{\bibinfo{person}{Ellie Harmon} {and} \bibinfo{person}{M~Six
  Silberman}.} \bibinfo{year}{2018}\natexlab{}.
\newblock \showarticletitle{Rating working conditions on digital labor
  platforms}.
\newblock \bibinfo{journal}{\emph{Computer Supported Cooperative Work (CSCW)}}
  \bibinfo{volume}{27}, \bibinfo{number}{3-6} (\bibinfo{year}{2018}),
  \bibinfo{pages}{1275--1324}.
\newblock


\bibitem[\protect\citeauthoryear{Hitlin}{Hitlin}{2016}]%
        {hitlin2016research}
\bibfield{author}{\bibinfo{person}{Paul Hitlin}.}
  \bibinfo{year}{2016}\natexlab{}.
\newblock \showarticletitle{Research in the crowdsourcing age, a case study}.
\newblock \bibinfo{journal}{\emph{Pew Research Center}}  \bibinfo{volume}{11}
  (\bibinfo{year}{2016}).
\newblock


\bibitem[\protect\citeauthoryear{Horowitz}{Horowitz}{2015}]%
        {costs_2019}
\bibfield{author}{\bibinfo{person}{Sara Horowitz}.}
  \bibinfo{year}{2015}\natexlab{}.
\newblock \bibinfo{title}{The Costs of Nonpayment}.
\newblock
\newblock
\urldef\tempurl%
\url{https://blog.freelancersunion.org/2015/12/10/costs-nonpayment/}
\showURL{%
Retrieved June 26, 2019 from \tempurl}


\bibitem[\protect\citeauthoryear{Horton}{Horton}{2011}]%
        {horton2011condition}
\bibfield{author}{\bibinfo{person}{John~J Horton}.}
  \bibinfo{year}{2011}\natexlab{}.
\newblock \showarticletitle{The condition of the Turking class: Are online
  employers fair and honest?}
\newblock \bibinfo{journal}{\emph{Economics Letters}} \bibinfo{volume}{111},
  \bibinfo{number}{1} (\bibinfo{year}{2011}), \bibinfo{pages}{10--12}.
\newblock


\bibitem[\protect\citeauthoryear{Horton and Chilton}{Horton and
  Chilton}{2010}]%
        {horton2010labor}
\bibfield{author}{\bibinfo{person}{John~Joseph Horton} {and}
  \bibinfo{person}{Lydia~B Chilton}.} \bibinfo{year}{2010}\natexlab{}.
\newblock \showarticletitle{The labor economics of paid crowdsourcing}. In
  \bibinfo{booktitle}{\emph{Proceedings of the 11th ACM conference on
  Electronic commerce}}. ACM, \bibinfo{pages}{209--218}.
\newblock


\bibitem[\protect\citeauthoryear{Huang, Chang, and Bigham}{Huang
  et~al\mbox{.}}{2018}]%
        {huang2018evorus}
\bibfield{author}{\bibinfo{person}{Ting-Hao~Kenneth Huang},
  \bibinfo{person}{Joseph~Chee Chang}, {and} \bibinfo{person}{Jeffrey~P
  Bigham}.} \bibinfo{year}{2018}\natexlab{}.
\newblock \showarticletitle{Evorus: A Crowd-powered Conversational Assistant
  Built to Automate Itself Over Time}. In \bibinfo{booktitle}{\emph{Proceedings
  of the 2018 CHI Conference on Human Factors in Computing Systems}}. ACM,
  \bibinfo{pages}{295}.
\newblock


\bibitem[\protect\citeauthoryear{(ILO)}{(ILO)}{2016}]%
        {international2016non}
\bibfield{author}{\bibinfo{person}{International Labour~Office (ILO)}.}
  \bibinfo{year}{2016}\natexlab{}.
\newblock \bibinfo{title}{Non-standard employment around the world:
  Understanding challenges, shaping prospects}.
\newblock
\newblock


\bibitem[\protect\citeauthoryear{Ipeirotis}{Ipeirotis}{2010}]%
        {ipeirotis2010mechanical}
\bibfield{author}{\bibinfo{person}{Panagiotis~G Ipeirotis}.}
  \bibinfo{year}{2010}\natexlab{}.
\newblock \showarticletitle{Mechanical Turk, low wages, and the market for
  lemons}.
\newblock \bibinfo{journal}{\emph{A Computer Scientist in a Business School}}
  \bibinfo{volume}{27} (\bibinfo{year}{2010}).
\newblock


\bibitem[\protect\citeauthoryear{Irani and Silberman}{Irani and
  Silberman}{2013}]%
        {irani2013turkopticon}
\bibfield{author}{\bibinfo{person}{Lilly~C Irani} {and} \bibinfo{person}{M
  Silberman}.} \bibinfo{year}{2013}\natexlab{}.
\newblock \showarticletitle{Turkopticon: Interrupting worker invisibility in
  amazon mechanical turk}. In \bibinfo{booktitle}{\emph{Proceedings of the
  SIGCHI conference on human factors in computing systems}}. ACM,
  \bibinfo{pages}{611--620}.
\newblock


\bibitem[\protect\citeauthoryear{Irani and Silberman}{Irani and
  Silberman}{2016}]%
        {irani2016stories}
\bibfield{author}{\bibinfo{person}{Lilly~C Irani} {and} \bibinfo{person}{M
  Silberman}.} \bibinfo{year}{2016}\natexlab{}.
\newblock \showarticletitle{Stories we tell about labor: Turkopticon and the
  trouble with design}. In \bibinfo{booktitle}{\emph{Proceedings of the 2016
  CHI conference on human factors in computing systems}}. ACM,
  \bibinfo{pages}{4573--4586}.
\newblock


\bibitem[\protect\citeauthoryear{Jarrahi and Sutherland}{Jarrahi and
  Sutherland}{2019}]%
        {jarrahi2019algorithmic}
\bibfield{author}{\bibinfo{person}{Mohammad~Hossein Jarrahi} {and}
  \bibinfo{person}{Will Sutherland}.} \bibinfo{year}{2019}\natexlab{}.
\newblock \showarticletitle{Algorithmic Management and Algorithmic
  Competencies: Understanding and Appropriating Algorithms in Gig work}. In
  \bibinfo{booktitle}{\emph{International Conference on Information}}.
  Springer, \bibinfo{pages}{578--589}.
\newblock


\bibitem[\protect\citeauthoryear{Jarrahi, Sutherland, Nelson, and
  Sawyer}{Jarrahi et~al\mbox{.}}{2019}]%
        {jarrahi2019platformic}
\bibfield{author}{\bibinfo{person}{Mohammad~Hossein Jarrahi},
  \bibinfo{person}{Will Sutherland}, \bibinfo{person}{Sarah~Beth Nelson}, {and}
  \bibinfo{person}{Steve Sawyer}.} \bibinfo{year}{2019}\natexlab{}.
\newblock \showarticletitle{Platformic Management, Boundary Resources for Gig
  Work, and Worker Autonomy}.
\newblock \bibinfo{journal}{\emph{Computer Supported Cooperative Work (CSCW)}}
  (\bibinfo{year}{2019}), \bibinfo{pages}{1--37}.
\newblock


\bibitem[\protect\citeauthoryear{Jiang and Li}{Jiang and Li}{1995}]%
        {jiang1995approximation}
\bibfield{author}{\bibinfo{person}{Tao Jiang} {and} \bibinfo{person}{Ming Li}.}
  \bibinfo{year}{1995}\natexlab{}.
\newblock \showarticletitle{On the approximation of shortest common
  supersequences and longest common subsequences}.
\newblock \bibinfo{journal}{\emph{SIAM J. Comput.}} \bibinfo{volume}{24},
  \bibinfo{number}{5} (\bibinfo{year}{1995}), \bibinfo{pages}{1122--1139}.
\newblock


\bibitem[\protect\citeauthoryear{Kaplan, Saito, Hara, and Bigham}{Kaplan
  et~al\mbox{.}}{2018}]%
        {kaplan2018striving}
\bibfield{author}{\bibinfo{person}{Toni Kaplan}, \bibinfo{person}{Susumu
  Saito}, \bibinfo{person}{Kotaro Hara}, {and} \bibinfo{person}{Jeffrey~P
  Bigham}.} \bibinfo{year}{2018}\natexlab{}.
\newblock \showarticletitle{Striving to Earn More: A Survey of Work Strategies
  and Tool Use Among Crowd Workers.}. In \bibinfo{booktitle}{\emph{HCOMP}}.
  \bibinfo{pages}{70--78}.
\newblock


\bibitem[\protect\citeauthoryear{Kasperczyk and Knickel}{Kasperczyk and
  Knickel}{1996}]%
        {kasperczyk1996analytic}
\bibfield{author}{\bibinfo{person}{N Kasperczyk} {and} \bibinfo{person}{K
  Knickel}.} \bibinfo{year}{1996}\natexlab{}.
\newblock \showarticletitle{The analytic hierarchy process (AHP)}.
\newblock \bibinfo{journal}{\emph{Retrieved from}} (\bibinfo{year}{1996}).
\newblock


\bibitem[\protect\citeauthoryear{K{\"a}ssi and Lehdonvirta}{K{\"a}ssi and
  Lehdonvirta}{2018}]%
        {kassi2018online}
\bibfield{author}{\bibinfo{person}{Otto K{\"a}ssi} {and} \bibinfo{person}{Vili
  Lehdonvirta}.} \bibinfo{year}{2018}\natexlab{}.
\newblock \showarticletitle{Online labour index: Measuring the online gig
  economy for policy and research}.
\newblock \bibinfo{journal}{\emph{Technological forecasting and social change}}
   \bibinfo{volume}{137} (\bibinfo{year}{2018}), \bibinfo{pages}{241--248}.
\newblock


\bibitem[\protect\citeauthoryear{Kasunic, Chiang, Kaufman, and Savage}{Kasunic
  et~al\mbox{.}}{2019}]%
        {kasunic2019crowd}
\bibfield{author}{\bibinfo{person}{Anna Kasunic}, \bibinfo{person}{Chun-Wei
  Chiang}, \bibinfo{person}{Geoff Kaufman}, {and} \bibinfo{person}{Saiph
  Savage}.} \bibinfo{year}{2019}\natexlab{}.
\newblock \showarticletitle{Crowd Work on a CV? Understanding How AMT Fits into
  Turkers' Career Goals and Professional Profiles}.
\newblock \bibinfo{journal}{\emph{arXiv preprint arXiv:1902.05361}}
  (\bibinfo{year}{2019}).
\newblock


\bibitem[\protect\citeauthoryear{Katz}{Katz}{2017}]%
        {KatzAmazon}
\bibfield{author}{\bibinfo{person}{Miranda Katz}.}
  \bibinfo{year}{2017}\natexlab{}.
\newblock \bibinfo{title}{Amazon Mechanical Turk Workers Have Had Enough}.
\newblock
\newblock
\urldef\tempurl%
\url{https://www.wired.com/story/amazons-turker-crowd-has-had-enough/}
\showURL{%
\tempurl}


\bibitem[\protect\citeauthoryear{Kaufmann, Schulze, and Veit}{Kaufmann
  et~al\mbox{.}}{2011}]%
        {kaufmann2011more}
\bibfield{author}{\bibinfo{person}{Nicolas Kaufmann}, \bibinfo{person}{Thimo
  Schulze}, {and} \bibinfo{person}{Daniel Veit}.}
  \bibinfo{year}{2011}\natexlab{}.
\newblock \showarticletitle{More than fun and money. Worker Motivation in
  Crowdsourcing-A Study on Mechanical Turk.}. In
  \bibinfo{booktitle}{\emph{AMCIS}}, Vol.~\bibinfo{volume}{11}. Detroit,
  Michigan, USA, \bibinfo{pages}{1--11}.
\newblock


\bibitem[\protect\citeauthoryear{Krafft, Young, Katell, Huang, and
  Bugingo}{Krafft et~al\mbox{.}}{2019}]%
        {krafft2019defining}
\bibfield{author}{\bibinfo{person}{PM Krafft}, \bibinfo{person}{Meg Young},
  \bibinfo{person}{Michael Katell}, \bibinfo{person}{Karen Huang}, {and}
  \bibinfo{person}{Ghislain Bugingo}.} \bibinfo{year}{2019}\natexlab{}.
\newblock \showarticletitle{Defining AI in Policy versus Practice}.
\newblock \bibinfo{journal}{\emph{arXiv preprint arXiv:1912.11095}}
  (\bibinfo{year}{2019}).
\newblock


\bibitem[\protect\citeauthoryear{Kuek, Paradi-Guilford, Fayomi, Imaizumi,
  Ipeirotis, Pina, and Singh}{Kuek et~al\mbox{.}}{2015}]%
        {kuek2015global}
\bibfield{author}{\bibinfo{person}{Siou~Chew Kuek}, \bibinfo{person}{Cecilia
  Paradi-Guilford}, \bibinfo{person}{Toks Fayomi}, \bibinfo{person}{Saori
  Imaizumi}, \bibinfo{person}{Panos Ipeirotis}, \bibinfo{person}{Patricia
  Pina}, {and} \bibinfo{person}{Manpreet Singh}.}
  \bibinfo{year}{2015}\natexlab{}.
\newblock \showarticletitle{The global opportunity in online outsourcing}.
\newblock  (\bibinfo{year}{2015}).
\newblock


\bibitem[\protect\citeauthoryear{Lampinen and Cheshire}{Lampinen and
  Cheshire}{2016}]%
        {lampinen2016hosting}
\bibfield{author}{\bibinfo{person}{Airi Lampinen} {and} \bibinfo{person}{Coye
  Cheshire}.} \bibinfo{year}{2016}\natexlab{}.
\newblock \showarticletitle{Hosting via Airbnb: Motivations and financial
  assurances in monetized network hospitality}. In
  \bibinfo{booktitle}{\emph{Proceedings of the 2016 CHI conference on human
  factors in computing systems}}. \bibinfo{pages}{1669--1680}.
\newblock


\bibitem[\protect\citeauthoryear{Lascau, Gould, Cox, Karmannaya, and
  Brumby}{Lascau et~al\mbox{.}}{2019}]%
        {lascau2019monotasking}
\bibfield{author}{\bibinfo{person}{Laura Lascau}, \bibinfo{person}{Sandy~JJ
  Gould}, \bibinfo{person}{Anna~L Cox}, \bibinfo{person}{Elizaveta Karmannaya},
  {and} \bibinfo{person}{Duncan~P Brumby}.} \bibinfo{year}{2019}\natexlab{}.
\newblock \showarticletitle{Monotasking or Multitasking: Designing for
  Crowdworkers' Preferences}. In \bibinfo{booktitle}{\emph{Proceedings of the
  2019 CHI Conference on Human Factors in Computing Systems}}.
  \bibinfo{pages}{1--14}.
\newblock


\bibitem[\protect\citeauthoryear{L{\'o}pez, Farzan, and Lin}{L{\'o}pez
  et~al\mbox{.}}{2017}]%
        {lopez2017behind}
\bibfield{author}{\bibinfo{person}{Claudia L{\'o}pez}, \bibinfo{person}{Rosta
  Farzan}, {and} \bibinfo{person}{Yu-Ru Lin}.} \bibinfo{year}{2017}\natexlab{}.
\newblock \showarticletitle{Behind the myths of citizen participation:
  Identifying sustainability factors of hyper-local information systems}.
\newblock \bibinfo{journal}{\emph{ACM Transactions on Internet Technology
  (TOIT)}} \bibinfo{volume}{18}, \bibinfo{number}{1} (\bibinfo{year}{2017}),
  \bibinfo{pages}{1--28}.
\newblock


\bibitem[\protect\citeauthoryear{Lynch, Blase, Wimms, Erikli, Benjafield,
  Kelly, and Willes}{Lynch et~al\mbox{.}}{2015}]%
        {lynch2015retrospective}
\bibfield{author}{\bibinfo{person}{S Lynch}, \bibinfo{person}{A Blase},
  \bibinfo{person}{A Wimms}, \bibinfo{person}{L Erikli}, \bibinfo{person}{A
  Benjafield}, \bibinfo{person}{C Kelly}, {and} \bibinfo{person}{L Willes}.}
  \bibinfo{year}{2015}\natexlab{}.
\newblock \showarticletitle{Retrospective descriptive study of CPAP adherence
  associated with use of the ResMed myAir application}.
\newblock \bibinfo{journal}{\emph{Available at:) ResMed Science Center, ResMed
  Ltd, Sydney (Australia)}} (\bibinfo{year}{2015}).
\newblock


\bibitem[\protect\citeauthoryear{Martin, Hanrahan, O'Neill, and Gupta}{Martin
  et~al\mbox{.}}{2014}]%
        {martin2014being}
\bibfield{author}{\bibinfo{person}{David Martin}, \bibinfo{person}{Benjamin~V
  Hanrahan}, \bibinfo{person}{Jacki O'Neill}, {and} \bibinfo{person}{Neha
  Gupta}.} \bibinfo{year}{2014}\natexlab{}.
\newblock \showarticletitle{Being a turker}. In
  \bibinfo{booktitle}{\emph{Proceedings of the 17th ACM conference on Computer
  supported cooperative work \& social computing}}. ACM,
  \bibinfo{pages}{224--235}.
\newblock


\bibitem[\protect\citeauthoryear{McInnis, Cosley, Nam, and Leshed}{McInnis
  et~al\mbox{.}}{2016}]%
        {mcinnis2016taking}
\bibfield{author}{\bibinfo{person}{Brian McInnis}, \bibinfo{person}{Dan
  Cosley}, \bibinfo{person}{Chaebong Nam}, {and} \bibinfo{person}{Gilly
  Leshed}.} \bibinfo{year}{2016}\natexlab{}.
\newblock \showarticletitle{Taking a HIT: Designing around rejection, mistrust,
  risk, and workers' experiences in Amazon Mechanical Turk}. In
  \bibinfo{booktitle}{\emph{Proceedings of the 2016 CHI conference on human
  factors in computing systems}}. ACM, \bibinfo{pages}{2271--2282}.
\newblock


\bibitem[\protect\citeauthoryear{Metall}{Metall}{2016}]%
        {metall2016frankfurt}
\bibfield{author}{\bibinfo{person}{IG Metall}.}
  \bibinfo{year}{2016}\natexlab{}.
\newblock \showarticletitle{Frankfurt paper on platform-based work—Proposals
  for platform operators, clients, policy makers, workers, and worker
  organizations}.
\newblock \bibinfo{journal}{\emph{IG Metall, Frankfurt}}
  (\bibinfo{year}{2016}).
\newblock


\bibitem[\protect\citeauthoryear{of~Labor}{of~Labor}{[n.d.]}]%
        {labor2019break}
\bibfield{author}{\bibinfo{person}{U.S.~Department of Labor}.}
  \bibinfo{year}{[n.d.]}\natexlab{}.
\newblock \bibinfo{title}{Breaks and Meal Periods}.
\newblock
\newblock
\urldef\tempurl%
\url{https://www.dol.gov/general/topic/workhours/breaks}
\showURL{%
\tempurl}


\bibitem[\protect\citeauthoryear{Park, Shoemark, and Morency}{Park
  et~al\mbox{.}}{2014}]%
        {park2014toward}
\bibfield{author}{\bibinfo{person}{Sunghyun Park}, \bibinfo{person}{Philippa
  Shoemark}, {and} \bibinfo{person}{Louis-Philippe Morency}.}
  \bibinfo{year}{2014}\natexlab{}.
\newblock \showarticletitle{Toward crowdsourcing micro-level behavior
  annotations: the challenges of interface, training, and generalization}. In
  \bibinfo{booktitle}{\emph{Proceedings of the 19th international conference on
  Intelligent User Interfaces}}. ACM, \bibinfo{pages}{37--46}.
\newblock


\bibitem[\protect\citeauthoryear{Payne, Bettman, and Luce}{Payne
  et~al\mbox{.}}{1996}]%
        {payne1996time}
\bibfield{author}{\bibinfo{person}{John~W Payne}, \bibinfo{person}{James~R
  Bettman}, {and} \bibinfo{person}{Mary~Frances Luce}.}
  \bibinfo{year}{1996}\natexlab{}.
\newblock \showarticletitle{When time is money: Decision behavior under
  opportunity-cost time pressure}.
\newblock \bibinfo{journal}{\emph{Organizational behavior and human decision
  processes}} \bibinfo{volume}{66}, \bibinfo{number}{2} (\bibinfo{year}{1996}),
  \bibinfo{pages}{131--152}.
\newblock


\bibitem[\protect\citeauthoryear{Persky and Robinson}{Persky and
  Robinson}{2017}]%
        {persky2017moving}
\bibfield{author}{\bibinfo{person}{Adam~M Persky} {and}
  \bibinfo{person}{Jennifer~D Robinson}.} \bibinfo{year}{2017}\natexlab{}.
\newblock \showarticletitle{Moving from novice to expertise and its
  implications for instruction}.
\newblock \bibinfo{journal}{\emph{American journal of pharmaceutical
  education}} \bibinfo{volume}{81}, \bibinfo{number}{9} (\bibinfo{year}{2017}),
  \bibinfo{pages}{6065}.
\newblock


\bibitem[\protect\citeauthoryear{Pritam}{Pritam}{2017}]%
        {pritam_2017}
\bibfield{author}{\bibinfo{person}{Nagrale Pritam}.}
  \bibinfo{year}{2017}\natexlab{}.
\newblock \bibinfo{title}{An Ultimate Guide to Making Money with Amazon mTurk}.
\newblock
\newblock
\urldef\tempurl%
\url{https://moneyconnexion.com/mturk-guide-to-make-money.htm}
\showURL{%
Retrieved June 26, 2019 from \tempurl}


\bibitem[\protect\citeauthoryear{Radford, Pilny, Reichelmann, Keegan, Welles,
  Hoye, Ognyanova, Meleis, and Lazer}{Radford et~al\mbox{.}}{2016}]%
        {radford2016volunteer}
\bibfield{author}{\bibinfo{person}{Jason Radford}, \bibinfo{person}{Andy
  Pilny}, \bibinfo{person}{Ashley Reichelmann}, \bibinfo{person}{Brian Keegan},
  \bibinfo{person}{Brooke~Foucault Welles}, \bibinfo{person}{Jefferson Hoye},
  \bibinfo{person}{Katherine Ognyanova}, \bibinfo{person}{Waleed Meleis}, {and}
  \bibinfo{person}{David Lazer}.} \bibinfo{year}{2016}\natexlab{}.
\newblock \showarticletitle{Volunteer science: An online laboratory for
  experiments in social psychology}.
\newblock \bibinfo{journal}{\emph{Social Psychology Quarterly}}
  \bibinfo{volume}{79}, \bibinfo{number}{4} (\bibinfo{year}{2016}),
  \bibinfo{pages}{376--396}.
\newblock


\bibitem[\protect\citeauthoryear{Ross, Irani, Silberman, Zaldivar, and
  Tomlinson}{Ross et~al\mbox{.}}{2010}]%
        {ross2010crowdworkers}
\bibfield{author}{\bibinfo{person}{Joel Ross}, \bibinfo{person}{Lilly Irani},
  \bibinfo{person}{M Silberman}, \bibinfo{person}{Andrew Zaldivar}, {and}
  \bibinfo{person}{Bill Tomlinson}.} \bibinfo{year}{2010}\natexlab{}.
\newblock \showarticletitle{Who are the crowdworkers?: shifting demographics in
  mechanical turk}. In \bibinfo{booktitle}{\emph{CHI'10 extended abstracts on
  Human factors in computing systems}}. ACM, \bibinfo{pages}{2863--2872}.
\newblock


\bibitem[\protect\citeauthoryear{Saito, Chiang, Savage, Nakano, Kobayashi, and
  Bigham}{Saito et~al\mbox{.}}{2019}]%
        {saito2019turkscanner}
\bibfield{author}{\bibinfo{person}{Susumu Saito}, \bibinfo{person}{Chun-Wei
  Chiang}, \bibinfo{person}{Saiph Savage}, \bibinfo{person}{Teppei Nakano},
  \bibinfo{person}{Tetsunori Kobayashi}, {and} \bibinfo{person}{Jeffrey
  Bigham}.} \bibinfo{year}{2019}\natexlab{}.
\newblock \showarticletitle{TurkScanner: Predicting the Hourly Wage of
  Microtasks}.
\newblock \bibinfo{journal}{\emph{arXiv preprint arXiv:1903.07032}}
  (\bibinfo{year}{2019}).
\newblock


\bibitem[\protect\citeauthoryear{Salehi, Irani, Bernstein, Alkhatib, Ogbe,
  Milland, et~al\mbox{.}}{Salehi et~al\mbox{.}}{2015}]%
        {salehi2015we}
\bibfield{author}{\bibinfo{person}{Niloufar Salehi}, \bibinfo{person}{Lilly~C
  Irani}, \bibinfo{person}{Michael~S Bernstein}, \bibinfo{person}{Ali
  Alkhatib}, \bibinfo{person}{Eva Ogbe}, \bibinfo{person}{Kristy Milland},
  {et~al\mbox{.}}} \bibinfo{year}{2015}\natexlab{}.
\newblock \showarticletitle{We are dynamo: Overcoming stalling and friction in
  collective action for crowd workers}. In
  \bibinfo{booktitle}{\emph{Proceedings of the 33rd annual ACM conference on
  human factors in computing systems}}. ACM, \bibinfo{pages}{1621--1630}.
\newblock


\bibitem[\protect\citeauthoryear{Sannon and Cosley}{Sannon and Cosley}{2019}]%
        {sannon2019privacy}
\bibfield{author}{\bibinfo{person}{Shruti Sannon} {and} \bibinfo{person}{Dan
  Cosley}.} \bibinfo{year}{2019}\natexlab{}.
\newblock \showarticletitle{Privacy, Power, and Invisible Labor on Amazon
  Mechanical Turk}.
\newblock  (\bibinfo{year}{2019}).
\newblock


\bibitem[\protect\citeauthoryear{Savage, Monroy-Hernandez, and
  H{\"o}llerer}{Savage et~al\mbox{.}}{2016}]%
        {savage2016botivist}
\bibfield{author}{\bibinfo{person}{Saiph Savage}, \bibinfo{person}{Andres
  Monroy-Hernandez}, {and} \bibinfo{person}{Tobias H{\"o}llerer}.}
  \bibinfo{year}{2016}\natexlab{}.
\newblock \showarticletitle{Botivist: Calling volunteers to action using online
  bots}. In \bibinfo{booktitle}{\emph{Proceedings of the 19th ACM Conference on
  Computer-Supported Cooperative Work \& Social Computing}}.
  \bibinfo{pages}{813--822}.
\newblock


\bibitem[\protect\citeauthoryear{Shadiev, Hwang, Yeh, Yang, Wang, Han, and
  Hsu}{Shadiev et~al\mbox{.}}{2014}]%
        {shadiev2014effects}
\bibfield{author}{\bibinfo{person}{Rustam Shadiev}, \bibinfo{person}{Wu-Yuin
  Hwang}, \bibinfo{person}{Shih-Ching Yeh}, \bibinfo{person}{Stephen~JH Yang},
  \bibinfo{person}{Jing-Liang Wang}, \bibinfo{person}{Lin Han}, {and}
  \bibinfo{person}{Guo-Liang Hsu}.} \bibinfo{year}{2014}\natexlab{}.
\newblock \showarticletitle{Effects of unidirectional vs. reciprocal teaching
  strategies on web-based computer programming learning}.
\newblock \bibinfo{journal}{\emph{Journal of educational computing research}}
  \bibinfo{volume}{50}, \bibinfo{number}{1} (\bibinfo{year}{2014}),
  \bibinfo{pages}{67--95}.
\newblock


\bibitem[\protect\citeauthoryear{Silberman and Metall}{Silberman and
  Metall}{2009}]%
        {silberman2009fifteen}
\bibfield{author}{\bibinfo{person}{M~Six Silberman} {and} \bibinfo{person}{IG
  Metall}.} \bibinfo{year}{2009}\natexlab{}.
\newblock \showarticletitle{Fifteen criteria for a fairer gig economy}.
\newblock \bibinfo{journal}{\emph{Democratization}} \bibinfo{volume}{61},
  \bibinfo{number}{4} (\bibinfo{year}{2009}), \bibinfo{pages}{589--622}.
\newblock


\bibitem[\protect\citeauthoryear{Silberman, Tomlinson, LaPlante, Ross, Irani,
  and Zaldivar}{Silberman et~al\mbox{.}}{2018}]%
        {silberman2018responsible}
\bibfield{author}{\bibinfo{person}{M~Six Silberman}, \bibinfo{person}{Bill
  Tomlinson}, \bibinfo{person}{Rochelle LaPlante}, \bibinfo{person}{Joel Ross},
  \bibinfo{person}{Lilly Irani}, {and} \bibinfo{person}{Andrew Zaldivar}.}
  \bibinfo{year}{2018}\natexlab{}.
\newblock \showarticletitle{Responsible research with crowds: pay crowdworkers
  at least minimum wage.}
\newblock \bibinfo{journal}{\emph{Commun. ACM}} \bibinfo{volume}{61},
  \bibinfo{number}{3} (\bibinfo{year}{2018}), \bibinfo{pages}{39--41}.
\newblock


\bibitem[\protect\citeauthoryear{Stiglitz}{Stiglitz}{2000}]%
        {stiglitz2000contributions}
\bibfield{author}{\bibinfo{person}{Joseph~E Stiglitz}.}
  \bibinfo{year}{2000}\natexlab{}.
\newblock \showarticletitle{The contributions of the economics of information
  to twentieth century economics}.
\newblock \bibinfo{journal}{\emph{The quarterly journal of economics}}
  \bibinfo{volume}{115}, \bibinfo{number}{4} (\bibinfo{year}{2000}),
  \bibinfo{pages}{1441--1478}.
\newblock


\bibitem[\protect\citeauthoryear{Strathern}{Strathern}{2000}]%
        {strathern2000tyranny}
\bibfield{author}{\bibinfo{person}{Marilyn Strathern}.}
  \bibinfo{year}{2000}\natexlab{}.
\newblock \showarticletitle{The tyranny of transparency}.
\newblock \bibinfo{journal}{\emph{British educational research journal}}
  \bibinfo{volume}{26}, \bibinfo{number}{3} (\bibinfo{year}{2000}),
  \bibinfo{pages}{309--321}.
\newblock


\bibitem[\protect\citeauthoryear{Suzuki, Salehi, Lam, Marroquin, and
  Bernstein}{Suzuki et~al\mbox{.}}{2016}]%
        {suzuki2016atelier}
\bibfield{author}{\bibinfo{person}{Ryo Suzuki}, \bibinfo{person}{Niloufar
  Salehi}, \bibinfo{person}{Michelle~S Lam}, \bibinfo{person}{Juan~C
  Marroquin}, {and} \bibinfo{person}{Michael~S Bernstein}.}
  \bibinfo{year}{2016}\natexlab{}.
\newblock \showarticletitle{Atelier: Repurposing expert crowdsourcing tasks as
  micro-internships}. In \bibinfo{booktitle}{\emph{Proceedings of the 2016 CHI
  Conference on Human Factors in Computing Systems}}. ACM,
  \bibinfo{pages}{2645--2656}.
\newblock


\bibitem[\protect\citeauthoryear{Vagias}{Vagias}{2006}]%
        {vagias2006likert}
\bibfield{author}{\bibinfo{person}{Wade~M Vagias}.}
  \bibinfo{year}{2006}\natexlab{}.
\newblock \showarticletitle{Likert-type scale response anchors. clemson
  international institute for tourism}.
\newblock \bibinfo{journal}{\emph{\& Research Development, Department of Parks,
  Recreation and Tourism Management, Clemson University}}
  (\bibinfo{year}{2006}).
\newblock


\bibitem[\protect\citeauthoryear{Vaidya and Kumar}{Vaidya and Kumar}{2006}]%
        {vaidya2006analytic}
\bibfield{author}{\bibinfo{person}{Omkarprasad~S Vaidya} {and}
  \bibinfo{person}{Sushil Kumar}.} \bibinfo{year}{2006}\natexlab{}.
\newblock \showarticletitle{Analytic hierarchy process: An overview of
  applications}.
\newblock \bibinfo{journal}{\emph{European Journal of operational research}}
  \bibinfo{volume}{169}, \bibinfo{number}{1} (\bibinfo{year}{2006}),
  \bibinfo{pages}{1--29}.
\newblock


\bibitem[\protect\citeauthoryear{Vakharia and Lease}{Vakharia and
  Lease}{2015}]%
        {vakharia2015beyond}
\bibfield{author}{\bibinfo{person}{Donna Vakharia} {and}
  \bibinfo{person}{Matthew Lease}.} \bibinfo{year}{2015}\natexlab{}.
\newblock \showarticletitle{Beyond Mechanical Turk: An analysis of paid crowd
  work platforms}.
\newblock \bibinfo{journal}{\emph{Proceedings of the iConference}}
  (\bibinfo{year}{2015}), \bibinfo{pages}{1--17}.
\newblock


\bibitem[\protect\citeauthoryear{Whiting, Hugh, and Bernstein}{Whiting
  et~al\mbox{.}}{2019}]%
        {whiting2019fair}
\bibfield{author}{\bibinfo{person}{Mark~E Whiting}, \bibinfo{person}{Grant
  Hugh}, {and} \bibinfo{person}{Michael~S Bernstein}.}
  \bibinfo{year}{2019}\natexlab{}.
\newblock \showarticletitle{Fair Work: Crowd Work Minimum Wage with One Line of
  Code}.
\newblock  (\bibinfo{year}{2019}).
\newblock


\bibitem[\protect\citeauthoryear{Williams, Mark, Milland, Lank, and
  Law}{Williams et~al\mbox{.}}{2019}]%
        {williams2019perpetual}
\bibfield{author}{\bibinfo{person}{Alex~C Williams}, \bibinfo{person}{Gloria
  Mark}, \bibinfo{person}{Kristy Milland}, \bibinfo{person}{Edward Lank}, {and}
  \bibinfo{person}{Edith Law}.} \bibinfo{year}{2019}\natexlab{}.
\newblock \showarticletitle{The Perpetual Work Life of Crowdworkers: How
  Tooling Practices Increase Fragmentation in Crowdwork}.
\newblock \bibinfo{journal}{\emph{Proceedings of the ACM on Human-Computer
  Interaction}} \bibinfo{volume}{3}, \bibinfo{number}{CSCW}
  (\bibinfo{year}{2019}), \bibinfo{pages}{1--28}.
\newblock


\bibitem[\protect\citeauthoryear{Yuen, King, and Leung}{Yuen
  et~al\mbox{.}}{2012}]%
        {yuen2012task}
\bibfield{author}{\bibinfo{person}{Man-Ching Yuen}, \bibinfo{person}{Irwin
  King}, {and} \bibinfo{person}{Kwong-Sak Leung}.}
  \bibinfo{year}{2012}\natexlab{}.
\newblock \showarticletitle{Task recommendation in crowdsourcing systems}. In
  \bibinfo{booktitle}{\emph{Proceedings of the first international workshop on
  crowdsourcing and data mining}}. ACM, \bibinfo{pages}{22--26}.
\newblock


\end{thebibliography}
\end{document}